\documentclass[12pt,preprint]{aastex}

\usepackage{natbib}
\begin{document}


\title{\ion{Si}{4} Column Densities Predicted from
  Non-Equilibrium Ionization Simulations of 
  Turbulent Mixing Layers and High-Velocity Clouds}


\author{Kyujin Kwak\altaffilmark{1}, 
Robin L. Shelton\altaffilmark{2}, and David B. Henley\altaffilmark{2}}
\affil{$^1$School of Natural Science, Ulsan National Institute of Science Institute (UNIST), 
50 UNIST-gil, Ulju-gun, Ulsan, 44919, Republic
of Korea}
\email{kkwak@unist.ac.kr}
\affil{$^2$Department of Physics and Astronomy, University of Georgia, 
Athens, GA 30602}
\email{rls@physast.uga.edu, dbh@physast.uga.edu}

\begin{abstract}

We present predictions of the \ion{Si}{4} ions in turbulent mixing layers 
(TMLs)
between hot and cool gas and in cool high velocity clouds (HVCs) that travel 
through a hot halo, 
complementing the \ion{C}{4}, \ion{N}{5}, and \ion{O}{6} predictions 
in Kwak \& Shelton (2010), Kwak et al. (2011), and Henley et al. (2012). 
%
%
We find that the \ion{Si}{4} ions are most abundant in 
regions where the hot and cool gases first begin to mix or
where the mixed gas has cooled significantly.
The predicted column densities of high velocity \ion{Si}{4} and the predicted
ratios of \ion{Si}{4} to \ion{C}{4} and \ion{O}{6} found on individual sightlines 
in our HVC simulations are in good agreement with 
observations of high velocity gas. Low velocity \ion{Si}{4} is also seen 
in the simulations,
as a result of decelerated gas in the case of the HVC simulations
and when looking along directions that pass perpendicular
to the direction of motion in the TML simulations. The ratios
of low velocity \ion{Si}{4} to \ion{C}{4} and \ion{O}{6} in the TML simulations
are in good agreement with those recorded for Milky Way
halo gas, while the ratio of \ion{Si}{4} to \ion{O}{6} from the decelerated
gas in the HVC simulations is lower than that observed at normal
velocity in the Milky Way halo. We attribute the shortfall of
normal velocity \ion{Si}{4} to not having modeled the effects of 
photoionization and, following Henley et al. (2012), consider a 
composite model that includes decelerated HVC gas, supernova 
remnants, galactic fountain gas, and the effect of photoionization.

\end{abstract}


\keywords{Galaxy: halo --- Hydrodynamics --- ISM: clouds ---
  Methods: numerical --- Ultraviolet: ISM}

\keywords{}



\section{Introduction}

Studies of the absorption lines in the ultraviolet (UV) spectra of
early type stars and active galactic nuclei (AGNs) have found large 
numbers of high ions, such as \ion{Si}{4}, \ion{C}{4},
\ion{N}{5}, and \ion{O}{6} in the Milky Way's disk 
\citep{Jenkins1978ApJa,Jenkins1978ApJb,
Cowieetal1981ApJ,SavageMassa1987ApJ,Savageetal2001ApJS,Bowenetal2008ApJS}
and halo \citep{SembachSavage1992ApJS,SavageSembach1994ApJ,
Sembachetal1997ApJ,Savageetal1997AJ,Savageetal2001ApJ,Sterlingetal2002ApJ,
Zsargoetal2003ApJ,Savageetal2003ApJS,Gangulyetal2005ApJS,Gangulyetal2006ApJ,
SavageWakker2009ApJ,Wakkeretal2012ApJ}.
The observations also reveal a population of high velocity 
($|v_{LSR}| \ga 90~\mbox{km}~\mbox{s}^{-1}$) high ions 
\citep{Sembachetal2003ApJS,Foxetal2004ApJ,Foxetal2005ApJ,Foxetal2010ApJ,
Collinsetal2004ApJ,Collinsetal2005ApJ,Collinsetal2007ApJ,
Gangulyetal2005ApJS,Shulletal2011ApJ}. Some of
these high-velocity high ions are observed along the same sightlines
where \ion{H}{1} high velocity clouds (HVCs) are observed, while
others appear where \ion{H}{1} HVCs have not been detected.

Transition temperature gas of $(0.5-10)\times 10^5$~K generally 
contains some high ions, but some \ion{Si}{4} and \ion{C}{4} can also 
exist at lower temperatures if sufficient ionizing radiation is incident upon the gas.
Although gas of this temperature is unstable due to 
its large radiative cooling rate, it may come from 
the interaction and/or transition between hotter ($T>10^6~\mbox{K}$) 
and cooler ($T < 10^4~\mbox{K}$) gases 
that are stable for longer periods of time. 


In previous studies, we examined the mixing between cool and hot 
gases. In the first study 
\citep[][hereafter KS10]{KwakShelton2010ApJ}, turbulent mixing of
cool and hot gas that slide past each other was
simulated in 2D planar (Cartesian) coordinates. Hereafter, 
we call this set of simulations `TML' (turbulent mixing layer) 
simulations. 
In the second study \citep[][hereafter KHS11]{Kwaketal2011ApJ}, 
we simulated cool spherical clouds falling through hot
ambient gas in 2D cylindrical coordinates 
(such simulations are called `HVC' simulations 
in this paper). 
In both sets of simulations, carried out with the FLASH hydrodynamics 
code \citep{Fryxelletal2000ApJS}, 
turbulent mixing is caused by shear instabilities that are instigated
by the velocity difference between cool and hot gas. 
High ions are abundant in the resulting mixed gas. 
Our modeling included 
the effect of radiative cooling, assuming 
collisional ionization equilibrium (CIE) cooling rates. 
We also modeled the ionization states of carbon,
nitrogen, and oxygen in the non-equilibrium fashion 
(i.e., the populations in all the ionization states of 
these elements were updated at each hydrodynamic time step).
Note that our TML and HVC simulations are only different in their 
geometries (planar in a 2D domain versus round in a 
cylindrically symmetric domain) while their modeling includes 
the same physical processes.

In KS10 and KHS11, we also compared 
our predictions of the column densities of \ion{C}{4}, 
\ion{N}{5}, and \ion{O}{6} (hereafter, 
N(\ion{C}{4}), N(\ion{N}{5}), and N(\ion{O}{6}), respectively) 
with the observed column densities of these ions in the Galactic halo 
(both low and high velocity ions)
and predictions from other models and found that 
using reasonable choices of model parameters, 
our model predictions are generally in good agreement with observations. 
Furthermore, in our more recent study 
\citep[][hereafter HKS12]{Henleyetal2012ApJ}, 
we estimated the amount of low-velocity \ion{C}{4}, 
\ion{N}{5}, and \ion{O}{6} predicted from the same set of simulations 
as KHS11. 
These low-velocity ions exist in the decelerated mixed material and 
we found that our model clouds could provide a significant fraction 
of the low-velocity \ion{O}{6} that is observed in the halo. 
However, the quantity of low-velocity \ion{C}{4} is 
not large enough to explain a significant fraction of 
the low-velocity \ion{C}{4} ions observed in the halo. 
Thus, HKS12 suggested that some of the low-velocity 
\ion{C}{4} is produced via photoionization, a phenomenon that the 
version of FLASH (v2.5) that we used does not model. Note that 
the latest version of FLASH (currently v4.2) still does not model 
photoionization.

In this paper, we present column densities of
\ion{Si}{4} (i.e., N(\ion{Si}{4})) 
that result from hydrodynamic interactions between hot and cool gas 
in the TML and HVC geometries.      
These calculations were done for the same physical set-ups as 
in KS10 and KHS11 and, like the results presented in those papers, 
do not include the effect of photoionization.
For this paper, 
we ran 13 additional FLASH runs (6 TML models like those in KS10 and 
7 HVC models like those in KHS11) that trace the ionization states of 
silicon. Note that the hydrodynamic evolution of the new FLASH runs 
in which the ion fractions of silicon ions were traced is identical to
those of the previous runs in which the ion fractions of 
the other three elements were traced\footnote{
The hydrodynamics calculated by the 
simulations are independent of the element whose ions we chose to trace 
because our cooling algorithm determines the cooling 
rate resulting from the dozen most abundant elements and the assumption that 
their ions are in CIE 
(see KS10 for some discussion of CIE versus NEI cooling rates).}. 
%
%
The results for N(\ion{Si}{4}) predicted from the additional simulations
are presented in a similar fashion as in KS10, KHS11, and HKS12. We 
compare our predictions of N(\ion{Si}{4}) with observations and
predictions from other types of models. 
The new results for N(\ion{Si}{4}) 
complement the previous results for N(\ion{C}{4}), N(\ion{N}{5}), and
N(\ion{O}{6}) that were presented in KS10, KHS11, and HKS12.

In HKS12, we considered a composite model in which the halo has 
four possible sources of low-velocity high ions, 
including our modeled HVCs, supernova remnants, galactic fountains, and 
photoionization. With this composite model, we tried to account for the quantities
of low-velocity \ion{C}{4}, \ion{N}{5}, and \ion{O}{6} 
observed in the Milky Way's halo. This composite model was designed to explain 
the full quantity of observed low-velocity \ion{O}{6} and we found that this 
model could also account for most of the observed low-velocity \ion{C}{4}. 
In this paper, we apply a similar analysis of the possible sources 
of low-velocity \ion{Si}{4} by using the same composite model as in HKS12. 
Note that among the four high ions observed in the UV band, 
\ion{Si}{4} can be produced by the lowest energy 
photons\footnote{The ionization potentials of \ion{Si}{3}, \ion{C}{3}, 
\ion{N}{4}, and \ion{O}{5}, which are the energies to further ionize these ions to the 
next level of ionization, are 33, 48, 77, and 114 eV, respectively. Note that 
the ionizing radiation field from 
hot (O-type) stars in the Galactic disk can provide photons with energies 
up to 54 eV, the \ion{He}{2} ionization edge, which are capable of photoionizing 
\ion{Si}{3} to \ion{Si}{4} and \ion{C}{3} to \ion{C}{4}, but are not capable 
of producing \ion{N}{5} or \ion{O}{6}. 
}. 
Hence, a significant fraction of \ion{Si}{4} 
observed in the Milky Way's halo could be produced via photoionization
as is the case with \ion{C}{4}. 
We find that an extra photoionization source which could increase only the number of 
low-velocity \ion{Si}{4} without affecting those of the other ions 
(especially \ion{C}{4}) would be required to explain the observed low-velocity 
\ion{Si}{4}. However, more detailed calculations that include more accurate 
modeling of the spectrum of ionizing photons will need to be 
carried out in order to explain the full amounts of low-velocity high ions 
in the halo. 
Note that photoionization can affect the amounts of high-velocity 
low ions such as \ion{C}{2}, \ion{C}{3}, \ion{Si}{2}, and \ion{Si}{3} in HVCs 
\citep{Foxetal2005ApJ} because escaping 
photons from the Galactic disk could reach even distant HVCs such as the Magellanic 
Stream \citep{BlandHawthornMaloney1999ApJL}.

In the next section, we will
briefly summarize the TML and HVC models from which the \ion{Si}{4}
column densities are predicted. The predicted 
\ion{Si}{4} column densities are presented 
in Sections \ref{TML_si4_S} and \ref{HVC_si4_S}, 
respectively. In Section \ref{compare_S}, we compare our \ion{Si}{4}
predictions with observations and with predictions made 
from other models. At the end of Section \ref{compare_S}, we will 
discuss the effect of photoionization on the \ion{Si}{4} ions in the 
Galactic halo by including our new \ion{Si}{4} results in the composite 
scenario like that considered in HKS12. 
Section \ref{summary_S} summarizes our results.

\section{Summary of NEI Simulations}\label{models_S}

The \ion{Si}{4} column densities that we present in this paper were 
calculated from new FLASH runs that modeled the same hydrodynamics 
as were modeled in KS10 and KHS11, with the only differences between 
the new and old FLASH runs being that silicon ionization levels 
were calculated in the new runs. 
In this section, we briefly summarize the basics of 
our previous two sets of simulations (TML from KS10 and HVC from KHS11).  
In the TML simulations, cool gas slid past hot gas. At the boundary, 
shear instigated instabilities that mixed the gas. An initial 
irregularity in the boundary between the two gases was used in 
order to seed the instabilities. 
In the reference model (TML Model A), the density and temperature 
of the cool and hot gas were chosen to match those of a 
cool cloud embedded in hot ambient gas in the Milky Way's halo 
($T_{cool}=10^3~\mbox{K}$,
$n_{cool}=0.1~\mbox{hydrogen atoms}~\mbox{cm}^{-3}$, 
$T_{hot}=10^6~\mbox{K}$, and
$n_{hot}=10^{-4}~\mbox{hydrogen atoms}~\mbox{cm}^{-3}$;
both cool and hot gas contain one helium atom per every 10 hydrogen atoms). 
The cool gas moved relative to the hot gas at a velocity of 
$v_x = 100~\mbox{km}~\mbox{s}^{-1}$. 
The simulation ran for 80~Myr. A layer of mixed gas developed 
between the cool and hot gas layers and it broadened over time. 
In KS10, we reported that high ions such as \ion{C}{4},
\ion{N}{5}, and \ion{O}{6} were abundant in the mixing
layer; here, we report that \ion{Si}{4} is also abundant. 
We refer readers to KS10 for more details regarding the 
TML models.




In the HVC simulations, spherically shaped clouds moved relative to 
the surrounding hot ambient medium. To be precise, the hot ambient medium 
flowed through the grid, past the initially stationary cloud. 
(Having the hot medium move rather than having the cloud move 
allowed us to track the evolution for longer periods of time.) 
The cloud's material ablated due to shear
instabilities and subsequently
mixed with the hot ambient medium, forming mixed gas where the high
ions became abundant. The reference model (HVC Model~B) had 
the ISM with the same density and temperature 
as the TML reference model (TML Model~A: $T_{ISM}=10^6~\mbox{K}$, and
$n_{ISM}=10^{-4}~\mbox{hydrogen atoms}~\mbox{cm}^{-3}$).  
The cloud's radius is approximately 150~pc and its 
density profile is shown in Figure 1 of KHS11. The cloud's 
temperature is determined from the cloud's density and 
the pressure balance with the ISM. Note that 
both cool and hot gas contain one helium atom per 10 hydrogen atoms 
and we assume that both hydrogen and helium are fully ionized 
regardless of the temperature during the simulation.
The hot medium moved relative to the cloud at a velocity of 
$100~\mbox{km}~\mbox{s}^{-1}$. 
More details regarding the HVC models can be found in KHS11.



For this paper, we ran 13 additional FLASH runs 
tracing the ionization states of silicon
(see Tables \ref{TMLmodels_T} and \ref{HVCmodels_T} 
for the model parameters of 6 TML models 
and 7 HVC models, respectively).
The FLASH NEI modules used the default abundance of silicon 
which is from \citet{LandiniMonsignoriFossi1990AAS}\footnote{
  The silicon abundance in \citet{LandiniMonsignoriFossi1990AAS}
  is $2.63 \times 10^{-5}$ (the ratio of silicon to hydrogen), 
  which is different from that in 
  \citet[][$3.31 \times 10^{-5}$]{Allen1973asqu.book}.} 
in these additional runs. However, the resulting column densities of
\ion{Si}{4} (and other high ions such as \ion{C}{4}, \ion{N}{5} and
\ion{O}{6}) can be linearly re-scaled if different choices of
abundances are desired, because the only effect of the chosen abundances 
on the results of our simulations is to change the total
number of ions of a given element\footnote{In principle, 
  the abundances could also affect the hydrodynamics of the
  simulations through the cooling rates. However, note that 
  in our current simulations 
  (both TML and HVC simulations), the cooling rates were approximated 
  with CIE cooling rates that were calculated from the 
  abundances of \cite{Allen1973asqu.book}. 
  Here we assumed that variations in abundances
  would not affect the CIE cooling rates significantly.}. 
In this paper, we rescaled the calculated \ion{Si}{4} column densities
using the interstellar silicon abundance of \citet{Wilmsetal2000ApJ}\footnote{
The abundances of carbon, nitrogen, oxygen, and silicon (i.e., the number per 
one hydrogen atom) are $2.40 \times 10^{-4}$, $7.59 \times 10^{-5}$, 
$4.90 \times 10^{-4}$, and $1.86 \times 10^{-5}$, respectively, 
in \citet{Wilmsetal2000ApJ}.}. 
Note that KS10 and KHS11 assumed \cite{Allen1973asqu.book} cosmic abundances, 
whereas HKS12 (our most recent paper on this topic) assumed \citet{Wilmsetal2000ApJ}
interstellar abundances.  
The \citet{Wilmsetal2000ApJ} abundances are in good
agreement with recently measured and commonly used solar photospheric
abundances \citep{Asplundetal2009ARAA}. 
For this reason, we also rescaled the column densities of other high ions
(\ion{C}{4}, \ion{N}{5}, and \ion{O}{6}) 
using the abundances of \citet{Wilmsetal2000ApJ}
for presentation in this paper.

\section{N(\ion{Si}{4}) from the TML Simulations}\label{TML_si4_S}

\begin{figure}
\epsscale{0.5}
\centering
\plotone{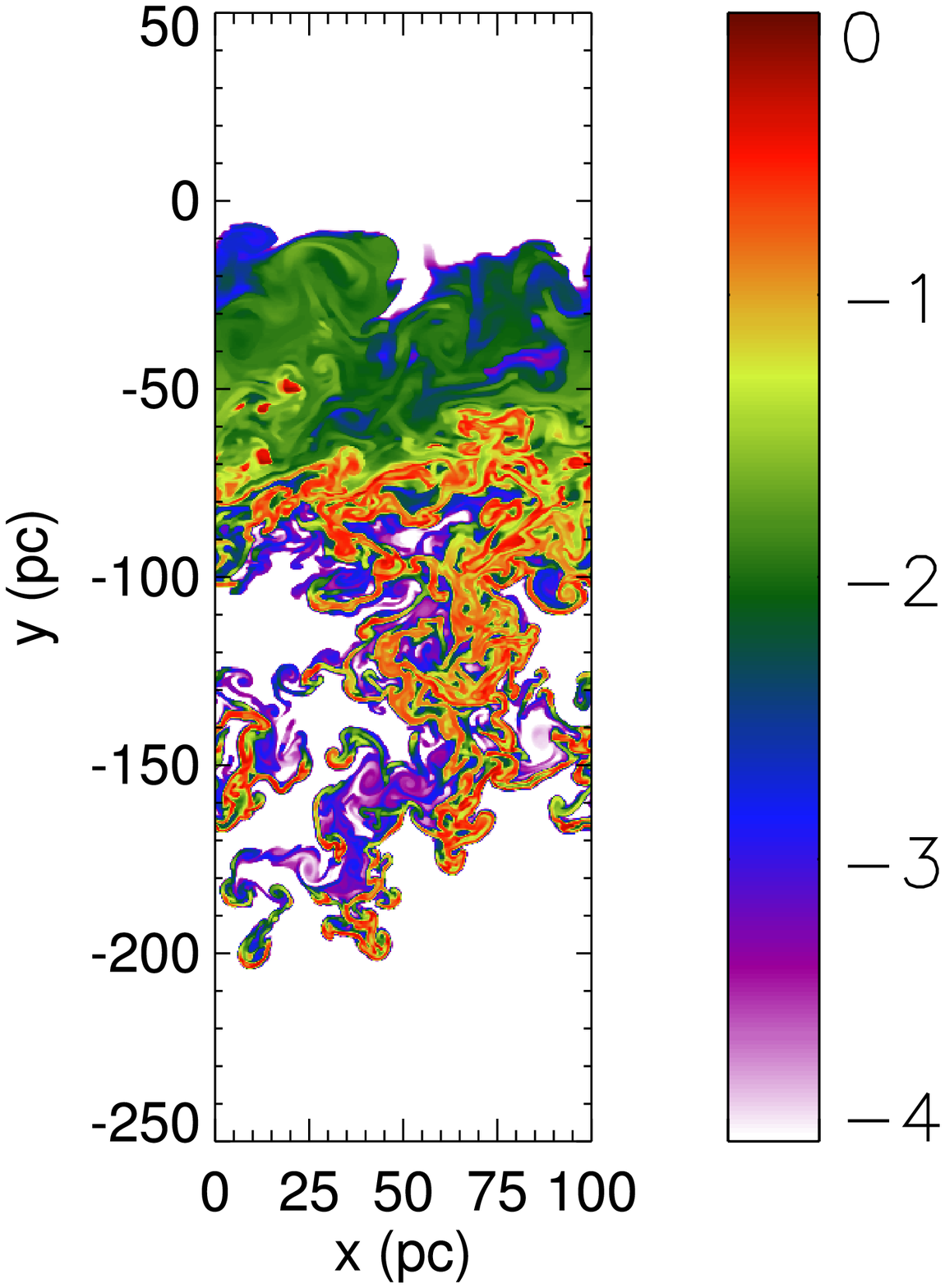}
\caption{The \ion{Si}{4} ion fraction (on a logarithmic scale) from TML Model~A 
  at t=80 Myr. 
  \label{TML_ModelA_Si04_80Myr_fig}}
\end{figure}

Here we discuss turbulent mixing in general and the \ion{Si}{4} ions 
modeled in our six new TML simulations in particular.
As the surface of a cool cloud slides across a reservoir of hot gas, 
shear instabilities (also known as Kelvin-Helmholtz instabilities) 
grow over time and as a result a mixing layer develops between 
the cold and hot regions. 
In this layer, material torn from the cool cloud mixes with hot gas 
and consequently rises in temperature and begins to ionize.  
Simultaneously, the portions of hot gas
that mixed with ablated cloud material begin to cool 
(due to mixing and due to radiation), and recombine. 
Owing to the slower rates at which recombination and ionization occur, 
the ionization levels of the gas throughout the mixing layer get
out of collisional ionization equilibrium and \ion{Si}{4} can be found 
throughout the mixed gas. Radiative cooling in the upper 
part of the mixing layer is particularly important, 
reducing its temperature to a few to ten times $10^3$~K much quicker than the 
ions can recombine. 
The non-equilibrium effects are very strong here, 
such that the fraction of silicon ions 
that have lost 3 electrons is much larger 
than that of CIE gas at the same temperature. 
Figure \ref{TML_ModelA_Si04_80Myr_fig} shows the fraction of Si ions that are 
in the \ion{Si}{4} level (the \ion{Si}{4} fraction) 
for a vertical slice through the domain at the end of the TML Model~A 
simulation (80 Myr). At this time, 
the depth of the entire mixing layer is about 200~pc, of which the region from a height 
of $z = -50$~pc to $z = 0$~pc has cooled to a few to ten times $10^3$~K and, 
as can be seen in the figure, still contains \ion{Si}{4}. Because shear instabilities 
still grow while the mixing layer develops, their features appear as the tendrils of 
the mixed gas in the region between $z \approx -100$~pc and $z \approx -200$~pc. 

The mixing layer is also rich in the other high ions, as can be seen in Figure 1 of KS10. 
A closer comparison between our Figure \ref{TML_ModelA_Si04_80Myr_fig} 
and Figure 1 of KS10, however, reveals that the evolution of \ion{Si}{4} resembles 
that of \ion{C}{4} more than those of the other ions. 
This is because the ionization potentials of \ion{Si}{3} and \ion{C}{3} 
are smaller and more similar to each other than those
of \ion{N}{4} and \ion{O}{5}.
Similar ionization patterns between \ion{C}{4} and \ion{Si}{4} within 
the temperature range of $10^4$~K to several times $10^5$~K can be understood 
as a result of similar electronic configurations: $1s^2 2s^1$ for \ion{C}{4} 
(the valence electron feels a net charge of $+4$ because the carbon nucleus’ $+6$ 
charge is screened by the two inner electrons) 
and $1s^2 2s^2 2p^6 3s^1$ for \ion{Si}{4} (similarly, the valence electron feels 
a net charge of $+4$ because the silicon nucleus’ $+14$ charge is screened 
by 10 inner electrons).

\begin{figure}
\centering
\plotone{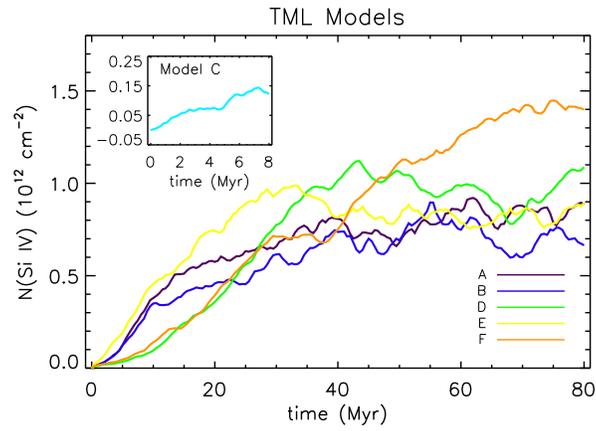}
\caption{Time evolution of the \ion{Si}{4} column densities for the TML
  models. The column densities at each epoch were obtained after
  averaging over many, evenly spaced, vertical sightlines through the domain 
  (256 sightlines for Models~A, B, D, E,
  and F and 512 sightlines for Model~B), i.e., sightlines 
  parallel to the $y$-axis at different $x$ locations 
  in Figure \ref{TML_ModelA_Si04_80Myr_fig}. 
  The inset is for Model C,
  which ran to t=8 Myr.
  \label{TML_si4_colm_fig}}
\end{figure}

\begin{figure}
\epsscale{1.0}
\centering
\plottwo{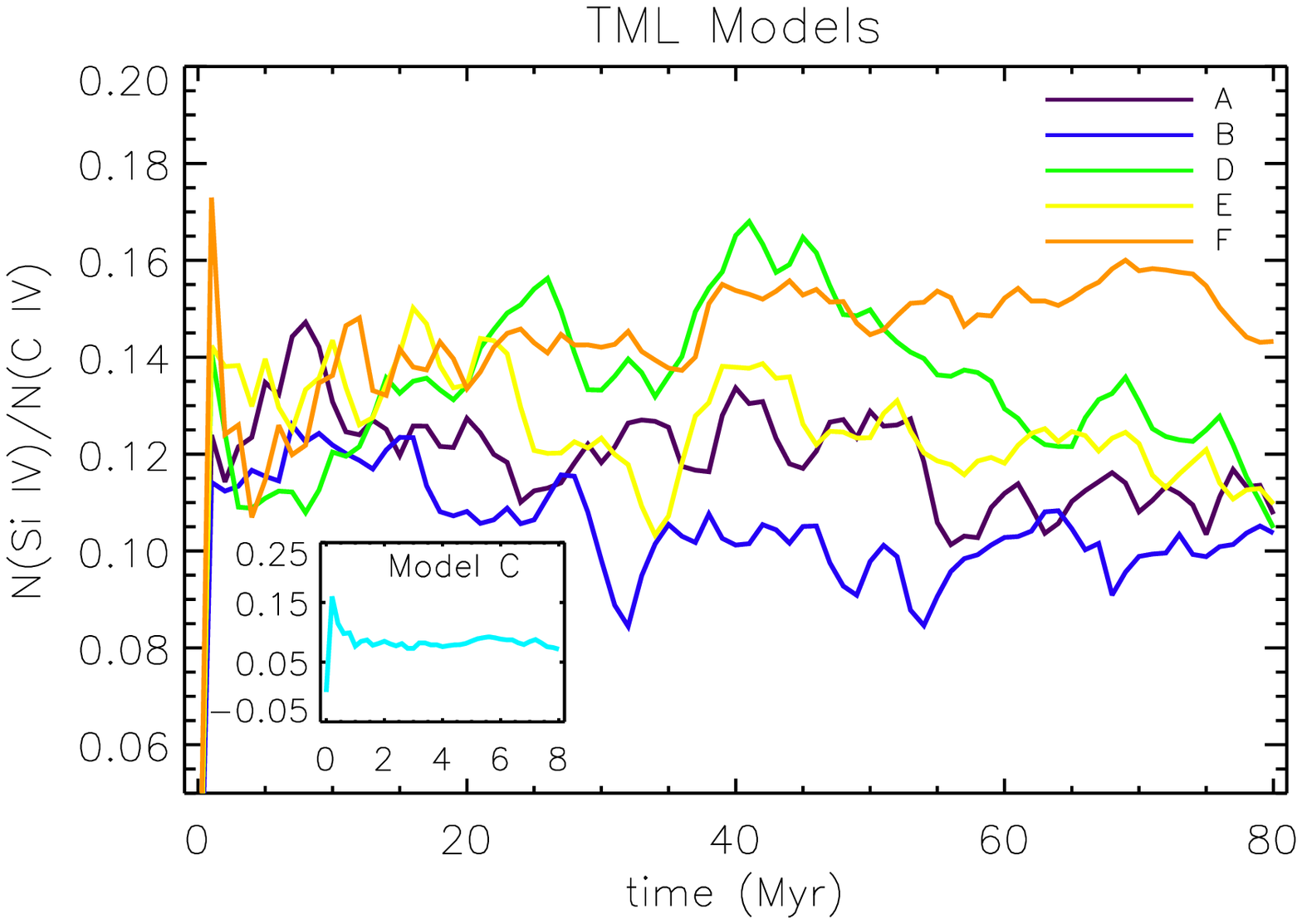}{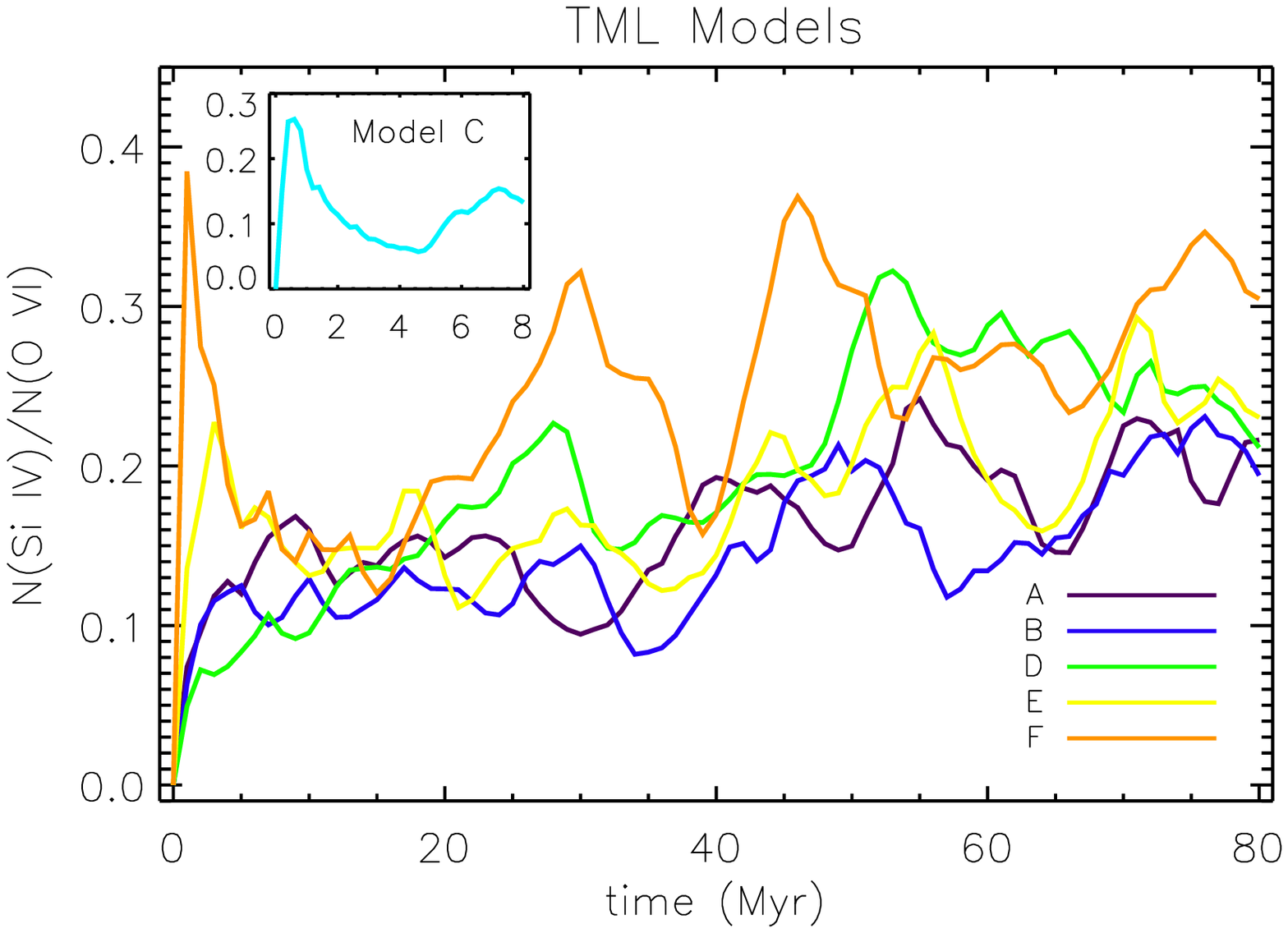}
\caption{Left and right panels, respectively, show 
  N(Si~IV)/N(C~IV) and N(Si~IV)/N(O~VI) 
  as a function of time from the TML models.  
  These ratios were calculated from column density averages that, 
  themselves were calculated from large numbers of vertical sightlines 
  through the domain 
  (256 sightlines for Models~A, B, D, E, and F and 512 sightlines 
  for Model~B). The inset is for Model C, which ran to t=8 Myr. 
  \label{TML_colm_ratio_fig}}
\end{figure}

The time evolution for N(\ion{Si}{4}) for TML Model~A and for the other models 
in the suite of TML simulations is shown in Figure \ref{TML_si4_colm_fig}. 
The \ion{Si}{4} column densities increase rapidly with time early in the simulations, 
as the mixing zone thickens, but less rapidly or not at all 
after the midpoints of the simulations. The exception to this trend is TML Model~F, 
which has a hotter ambient temperature than the other models and so its mixing zone 
requires more time to mature than in the other models. In general, the time evolution of 
N(\ion{Si}{4}) is similar to those of N(\ion{C}{4}), N(\ion{N}{5}), and N(\ion{O}{6}), 
which are shown in Figures 2 and 6 of KS 10. However, the predicted column densities 
of \ion{Si}{4} are smaller than those of the other ions because the gas has a lower 
abundance of silicon than of carbon, nitrogen, or oxygen.
The column densities of \ion{Si}{4} relative to those of \ion{C}{4} and \ion{O}{6} 
are shown in the ratio plots of Figure \ref{TML_colm_ratio_fig}. 
As mentioned above, N(\ion{Si}{4}) and
N(\ion{C}{4}) evolve similarly, thus the N(\ion{Si}{4})/N(\ion{C}{4}) ratio 
remains roughly constant over time for most of the models. 
We calculate the mean, median, standard deviation, minimum, and
maximum for N(\ion{Si}{4})/N(\ion{C}{4}) and
N(\ion{Si}{4})/N(\ion{O}{6}) within a range of time and tabulate the
results in Table \ref{TLM_colm_ratio_T}. 

\section{N(\ion{Si}{4}) from HVC Simulations}\label{HVC_si4_S}

\begin{figure}

\centering

\includegraphics[scale=0.25]{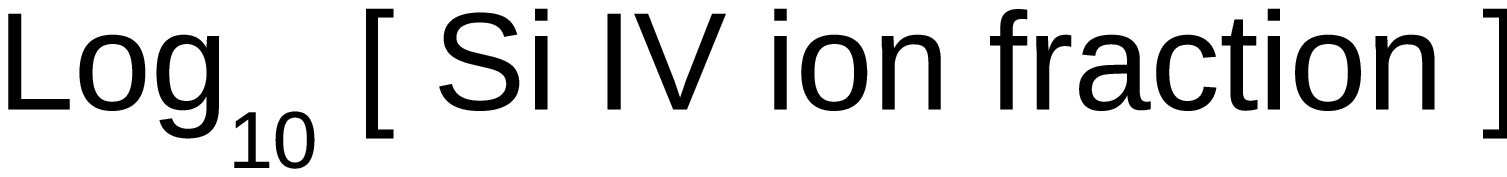} \\
\includegraphics[scale=0.18]{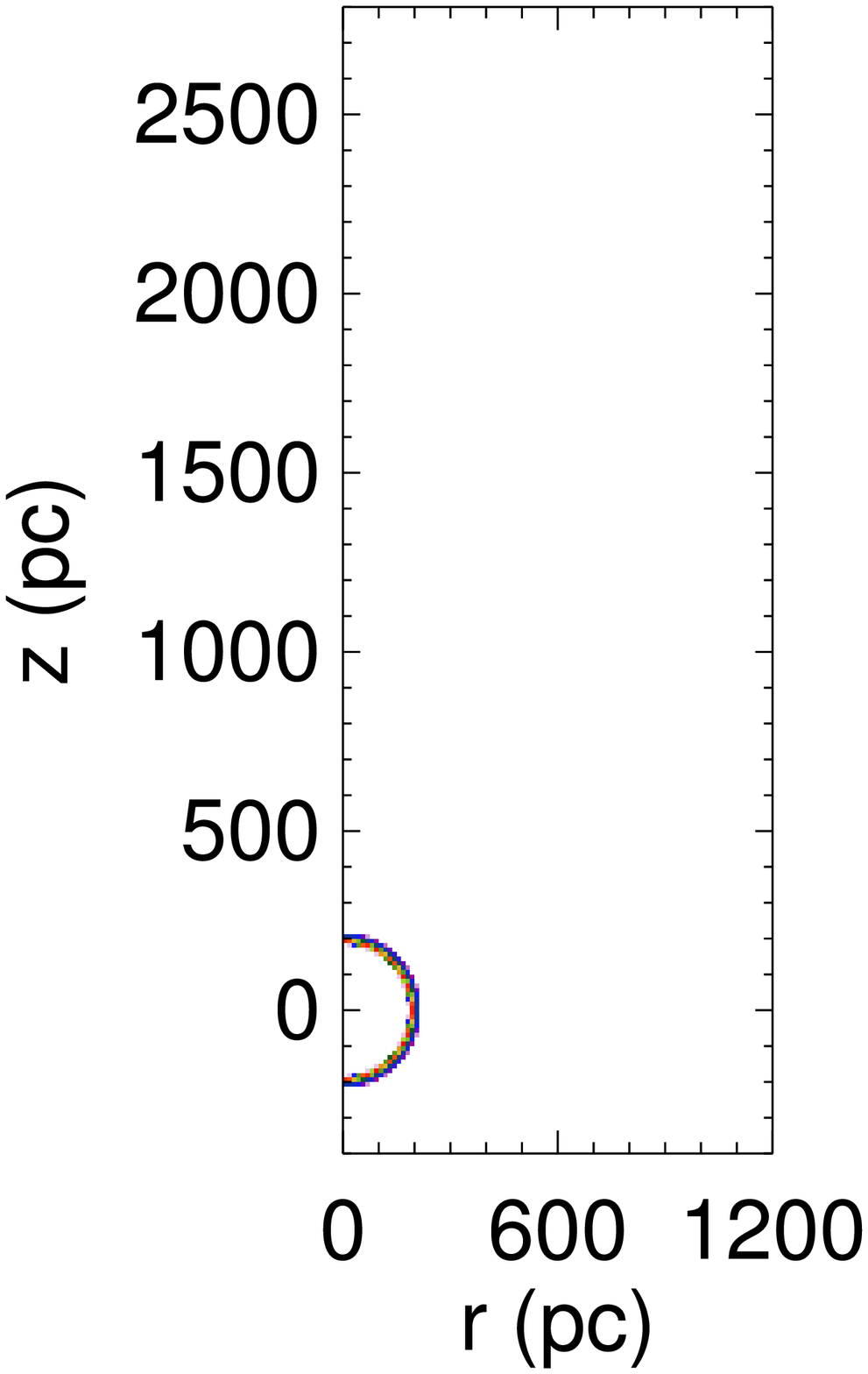}
\includegraphics[scale=0.18]{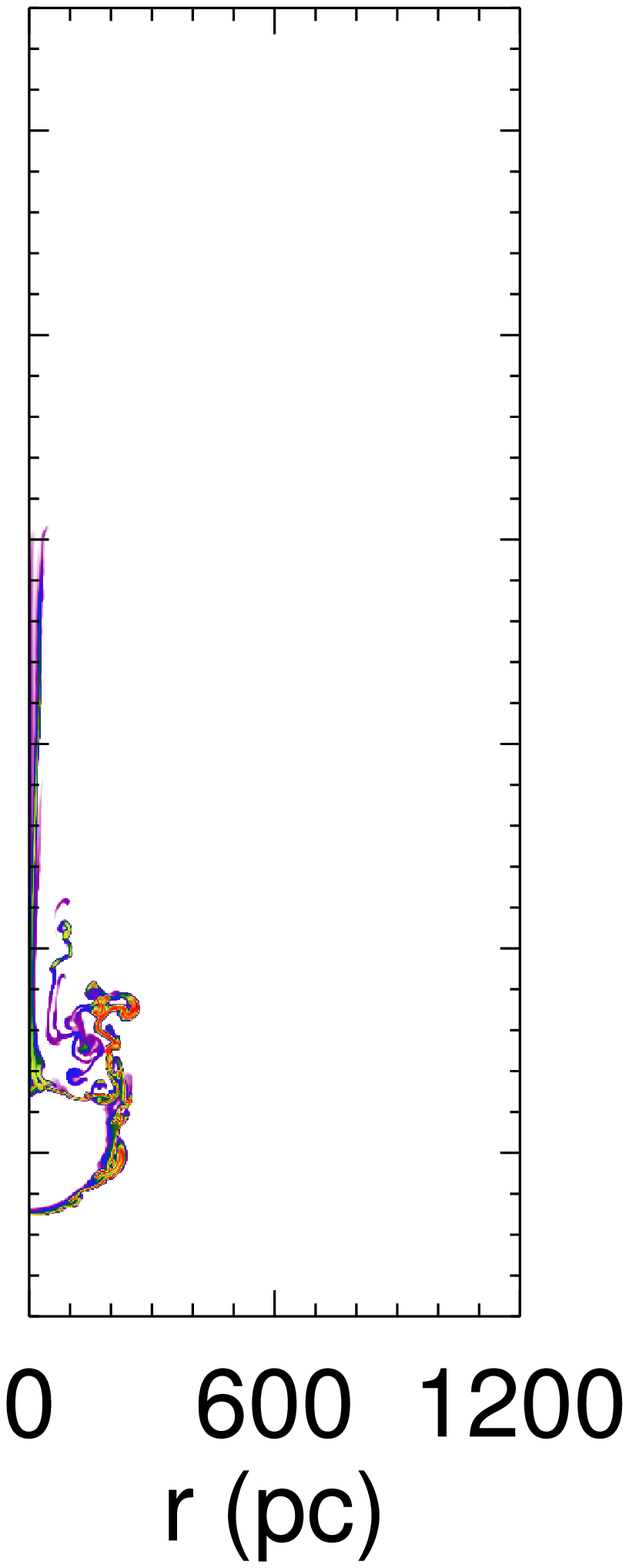}
\includegraphics[scale=0.18]{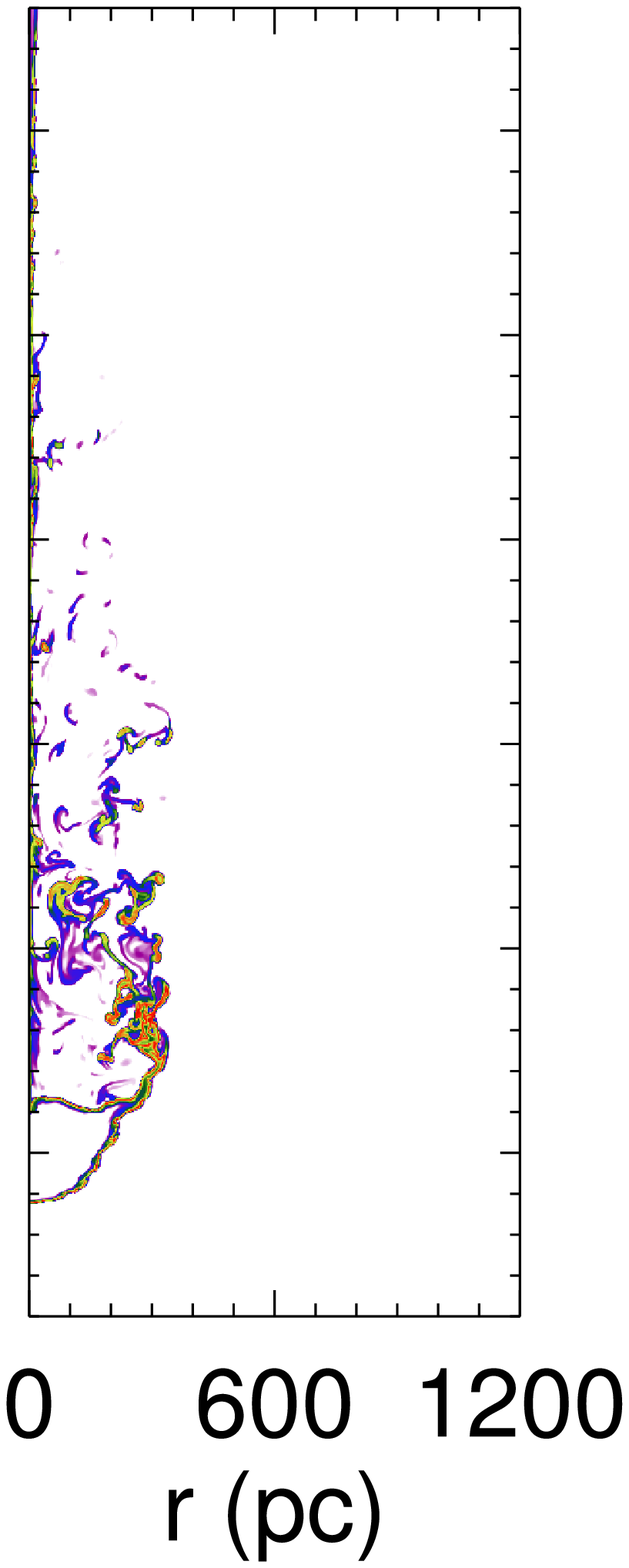}
\includegraphics[scale=0.18]{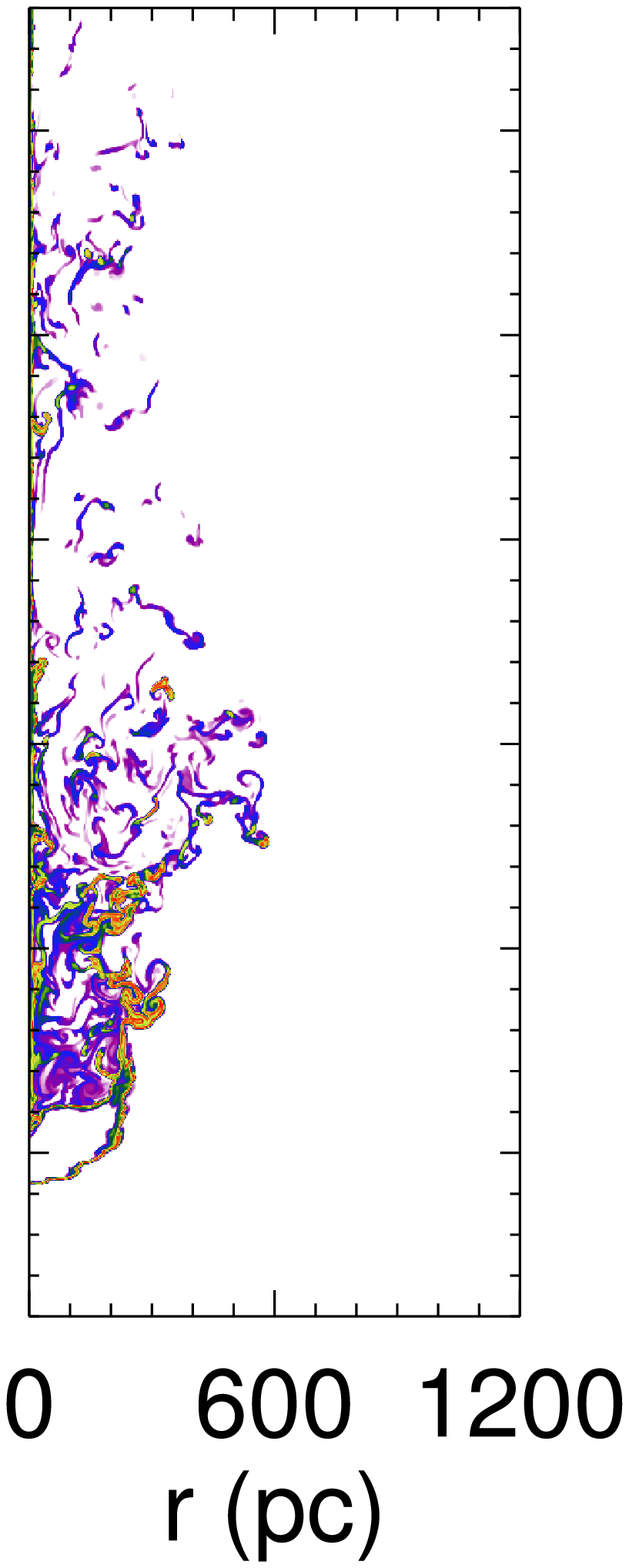}
\includegraphics[scale=0.18]{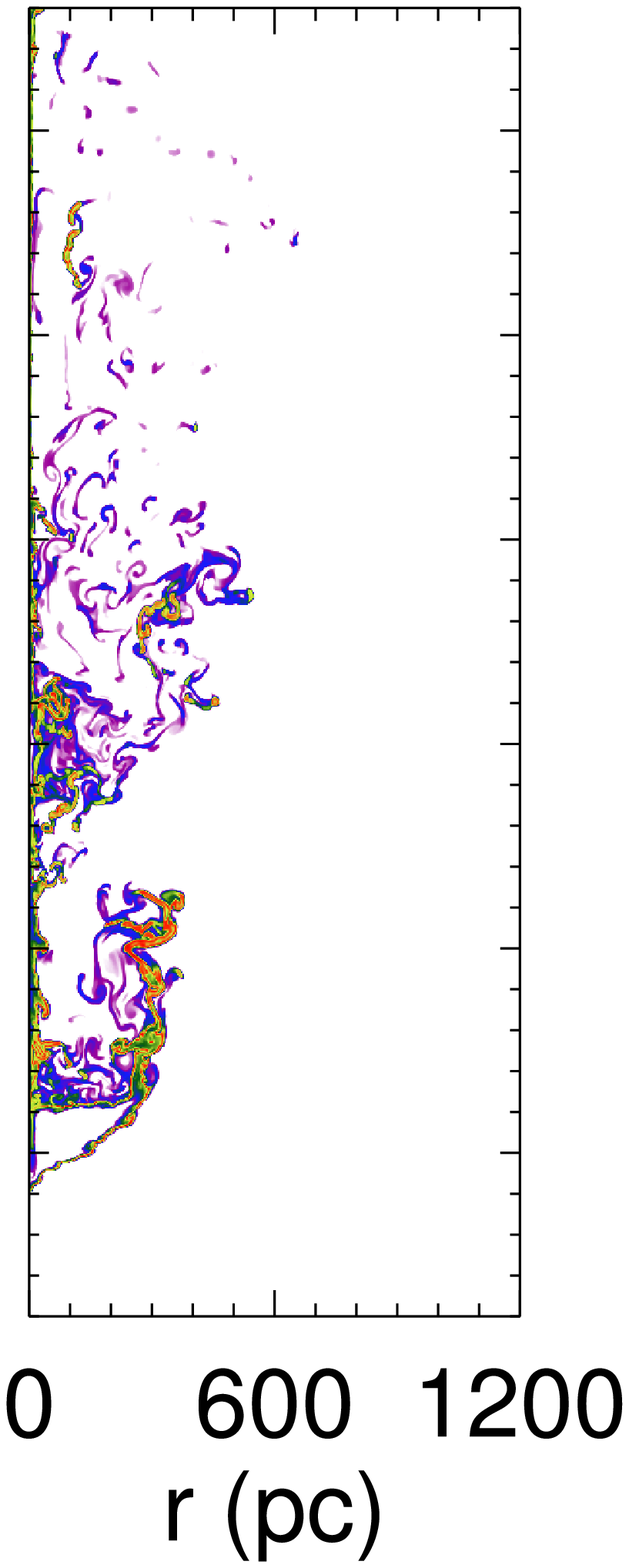}
\includegraphics[scale=0.18]{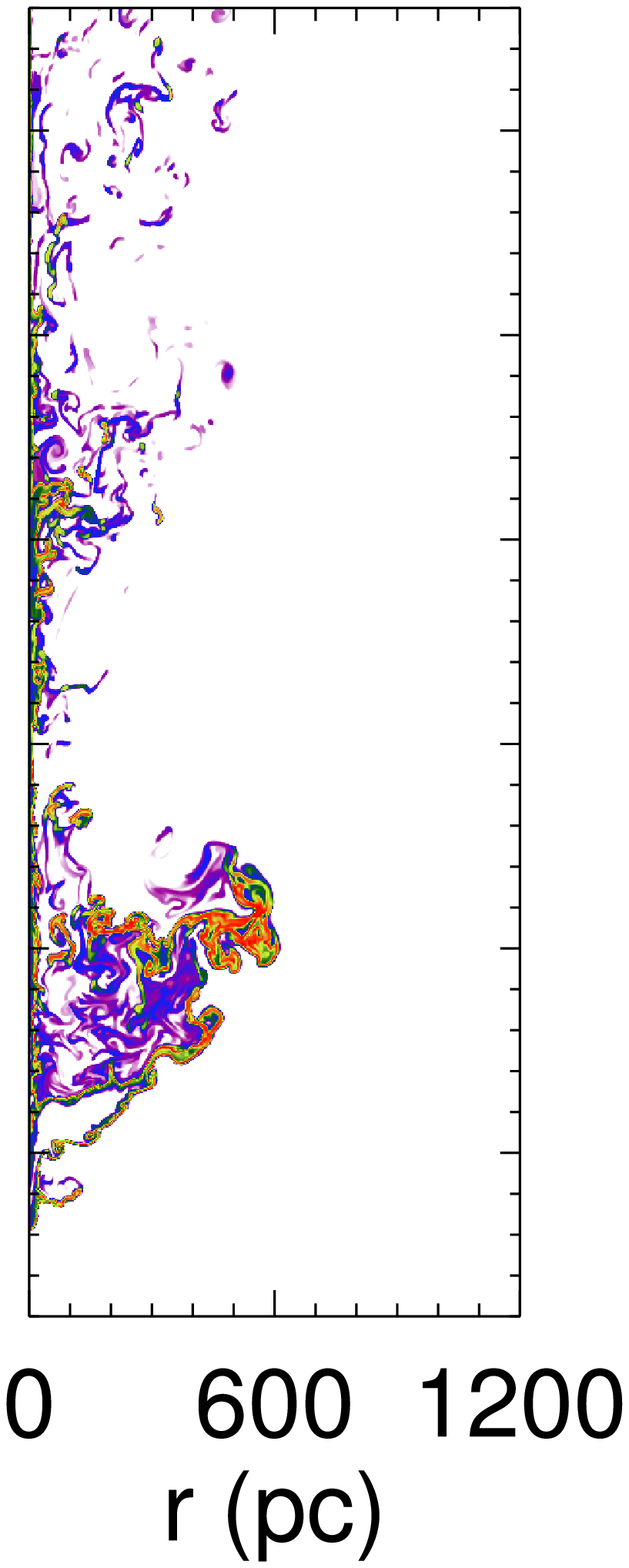}
\includegraphics[scale=0.18]{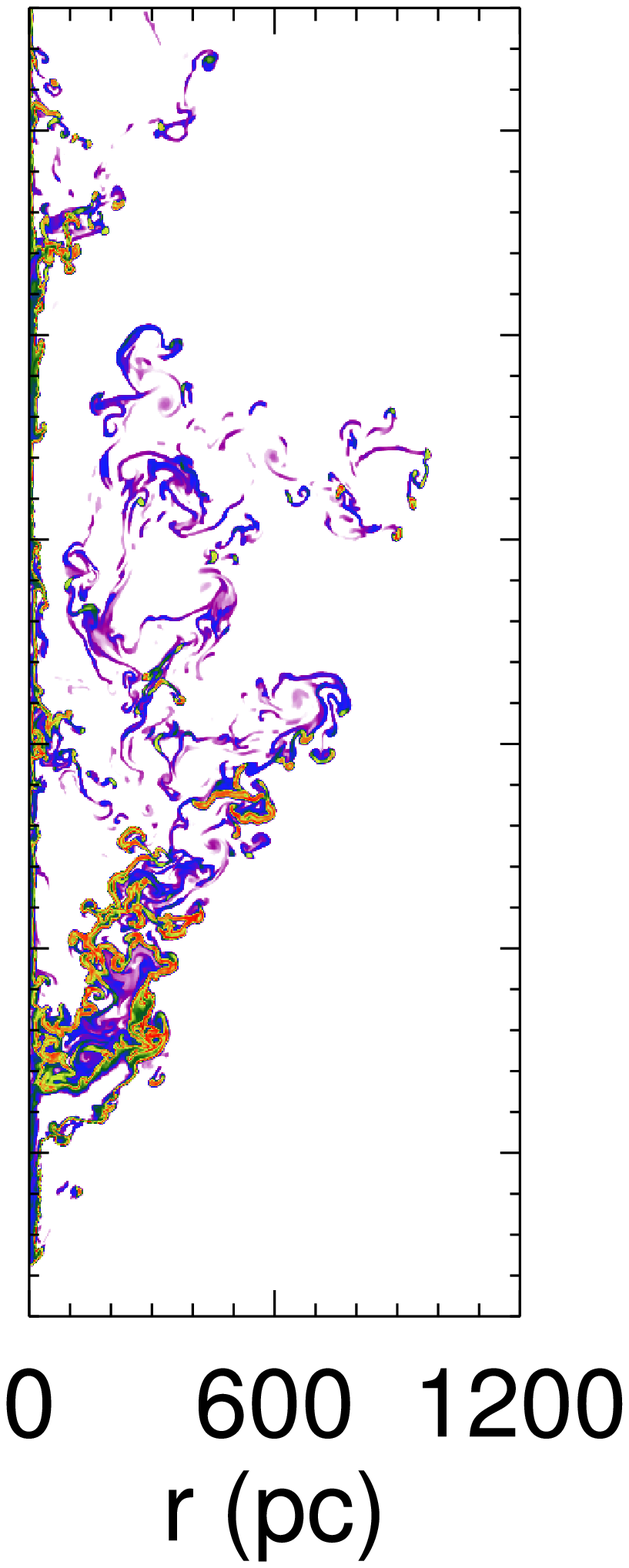}
\includegraphics[scale=0.18]{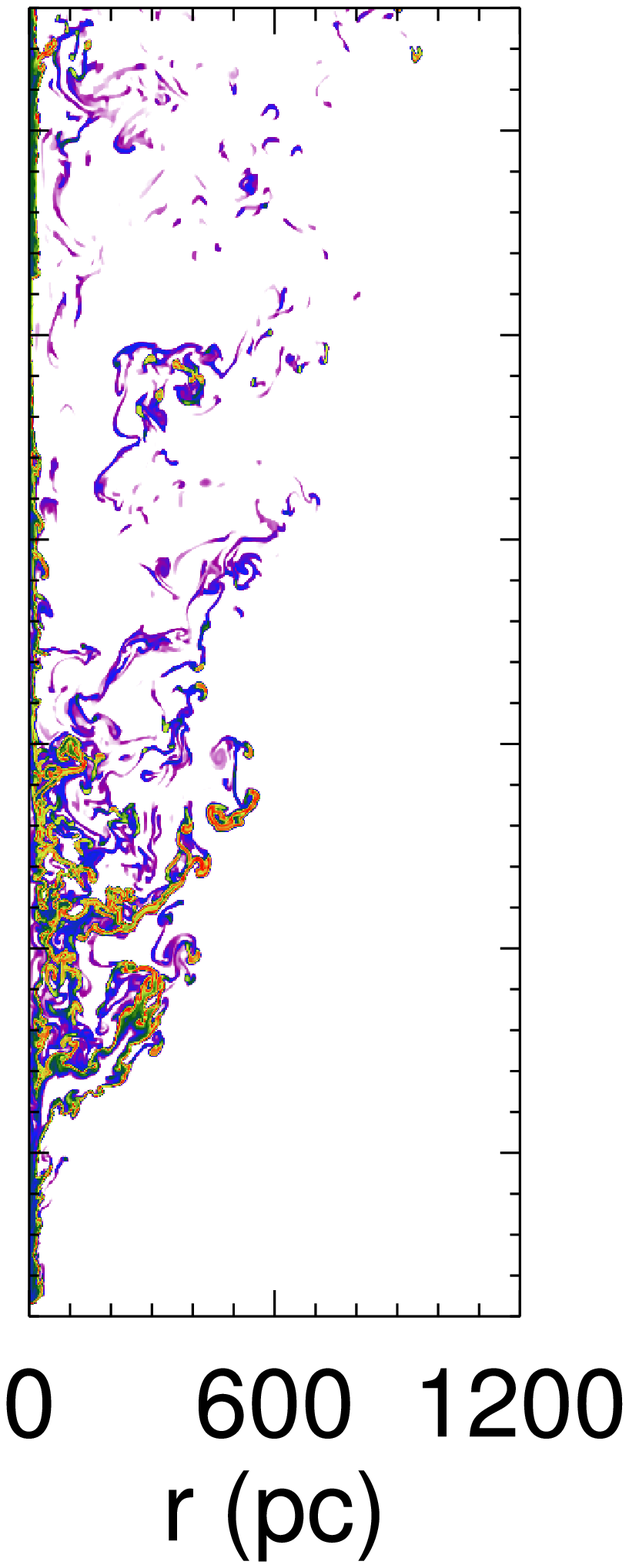}
\includegraphics[scale=0.18]{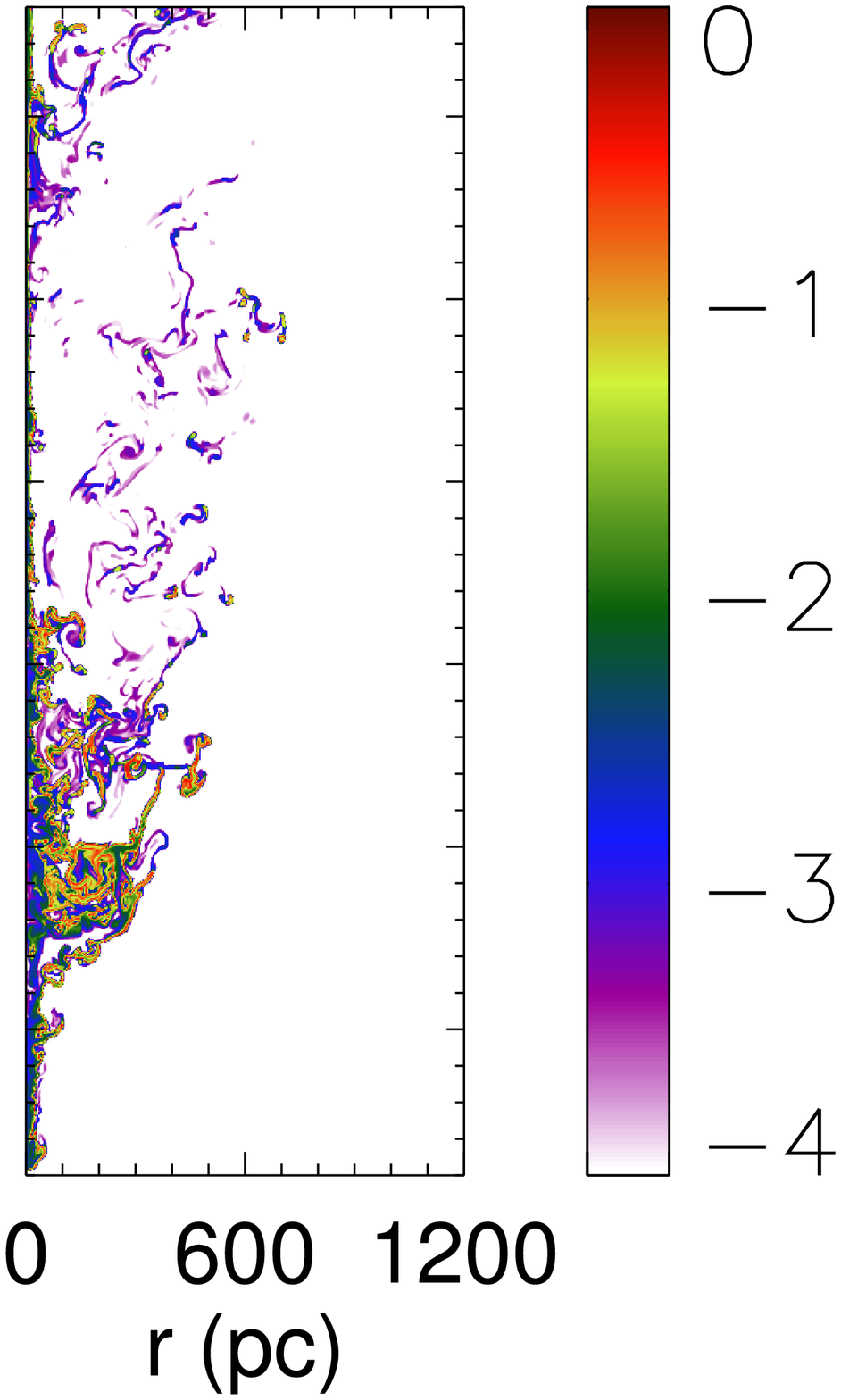}
\caption{\ion{Si}{4} ion fraction from HVC Model~B with logarithmic color scale.
  From left to right, each panel is plotted at 15 Myr intervals from $t=0$ to $t=120$~Myr. 
  The simulations were carried out in 2D cylindrical coordinates, and only half of 
  the spherical cloud is shown along a plane that cuts through the center of the cloud.
  \label{HVC_ModelB_fig}}
\end{figure}

In our HVC simulations, 
the initial spherical cloud loses its cool material due to shear instabilities 
that are developed across the boundary between the cloud and the ISM. 
As time goes by, the cold material ablated from the cloud mixes with 
the hot ambient gas and the mixed gas reaches intermediate temperatures 
due to both mixing and cooling. 
Figure \ref{HVC_ModelB_fig} shows the time 
evolution of the \ion{Si}{4} ion fraction obtained from HVC Model~B 
which is a reference model of our HVC simulations.
As in the TML simulations, \ion{Si}{4} becomes abundant in the intermediate temperature 
gas that results from the mixing of hot ambient gas with cold gas ablated from the 
cloud (see Figure 2 of KHS11 for the time evolution of hydrogen number density, temperature, 
$v_z$, and ion fractions of other ions, \ion{C}{4}, \ion{N}{5}, and \ion{O}{6}).
Mixed gas forms a trail
behind the cloud, with the mixed gas nearest to the cloud having approximately the same 
speed as the cloud and the mixed gas furthest from the cloud having slowed to approximately
the rest frame of the environment. Thus, in our HVC 
simulations, high ions including \ion{Si}{4} contained in the mixed gas are identified as 
either low- or high-velocity ions depending on the speed of 
the mixed gas relative to an observer in the rest frame 
of the environmental gas. Following the same convention as 
in KHS11 and HKS12, we distinguished the low- and 
high-velocity \ion{Si}{4} by using 
the same velocity criterion ($80~\mbox{km}~\mbox{s}^{-1}$ for HVC Models~A, B, 
F, and G and $100~\mbox{km}~\mbox{s}^{-1}$ for HVC Models~C, D, 
and E)\footnote{Our HVC simulations were conducted
with the ambient gas flowing upwards
($+z$ direction) while the initial cloud 
was stationary. However, in both KHS11 and HKS12, we reported 
the velocities in the observer's reference frame, 
where the observer is assumed to be 
located below the domain and moves with the same velocity 
as the initial ambient gas. In the 
observer's frame, the initial cloud falls downwards ($-z$ direction). 
We refer to material that travels downwards (in the observer's reference 
frame) with a velocity that is significantly less than the designated HVC velocity 
as low velocity material and material that travels downwards at a faster 
velocity as high velocity material.}.

\begin{figure}
  \epsscale{0.32}
  \centering
  \plotone{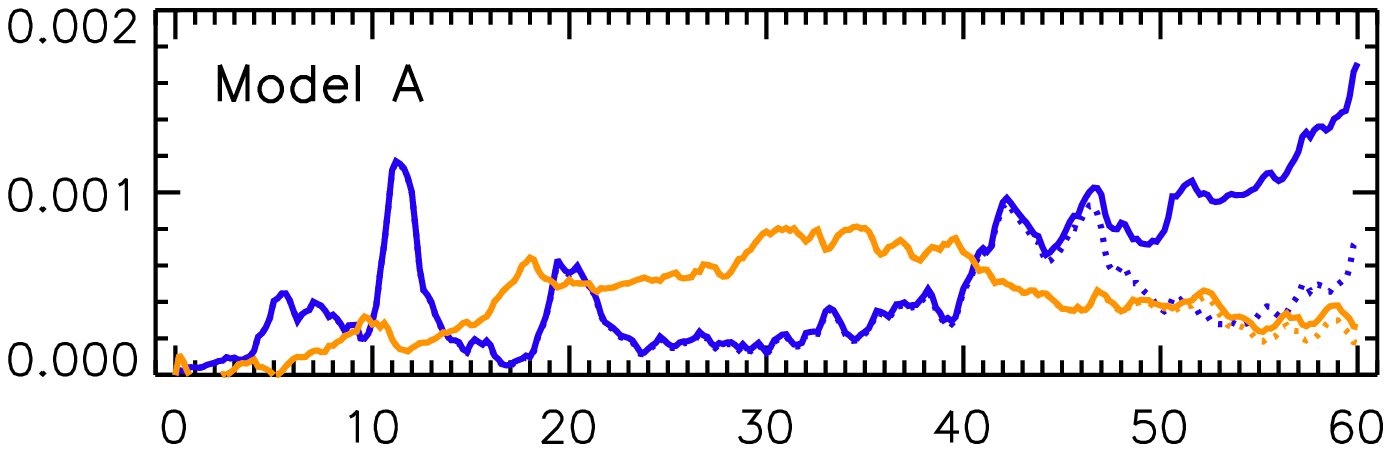} \\
  \plotone{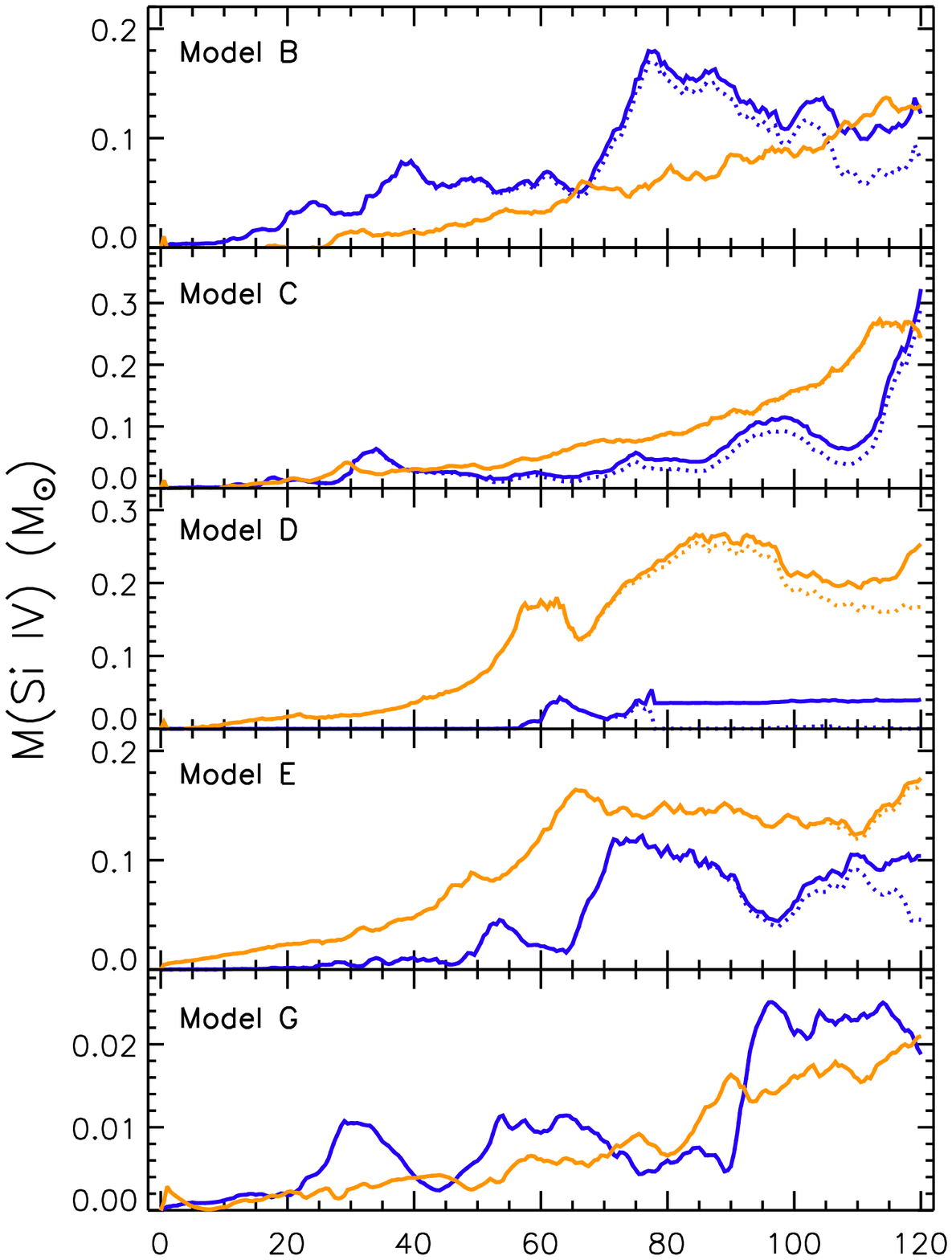} \\
  \plotone{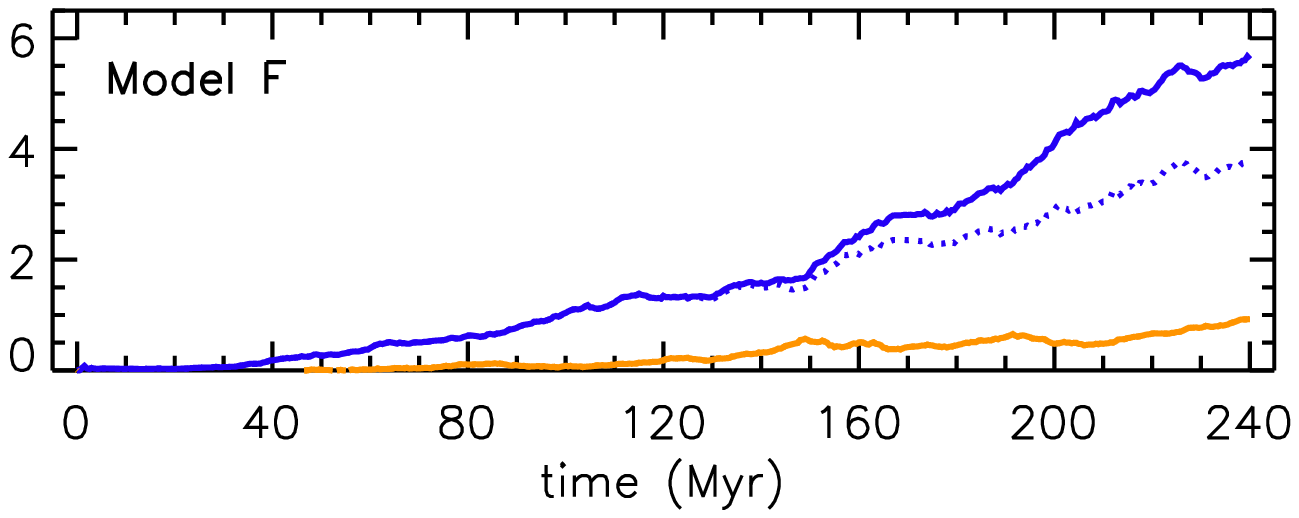}
  \caption{Mass (in solar masses) of Si~IV 
    as a function of time from seven HVC models.
    In each panel, the blue and orange lines correspond to 
    the masses of low- and high-velocity Si~IV, respectively.
    The dotted lines trace the Si~IV masses that are in the 
    domain as a function of time while 
    the solid lines track the sum of the mass of Si~IV ions 
    that are in the domain plus the mass of Si~IV ions that 
    have escaped from the domain by flowing through 
    the upper boundary of the domain. In most panels, the orange 
    dotted lines are overlapped by the orange solid lines and so 
    are not readily apparent. 
    \label{si4_mass_evol_fig}}
\end{figure}

We present first the evolution of the amounts of low velocity and high velocity 
\ion{Si}{4} in the mixed gas in the domain for each of our HVC models. 
Figure \ref{si4_mass_evol_fig} shows 
the mass of \ion{Si}{4} as a function of time and its evolution generally 
looks more similar to that 
of \ion{C}{4} (Figure 3 of HKS12) than to those of \ion{N}{5} and 
\ion{O}{6} (Figures 4 and 5 of HKS12, respectively). 
In particular, in most models, the tallies of low-velocity 
\ion{C}{4} and \ion{Si}{4} 
that include escaped ions 
(i.e., ions that have escaped from the domain by flowing through 
the upper boundary of the domain) 
tend to be comparable to or smaller than those for the high-velocity 
\ion{C}{4} and \ion{Si}{4}, in contrast with the case
of \ion{O}{6}, in which there tends to be more low velocity material 
than high velocity material at any given time. 
The resemblance in the mass evolution 
between \ion{C}{4} and \ion{Si}{4}, especially in comparison 
with \ion{O}{6}, can also be understood as being a result of 
carbon and silicon having similar ionization potentials.

\begin{figure}
  \epsscale{1.0}
  \centering
  \plottwo{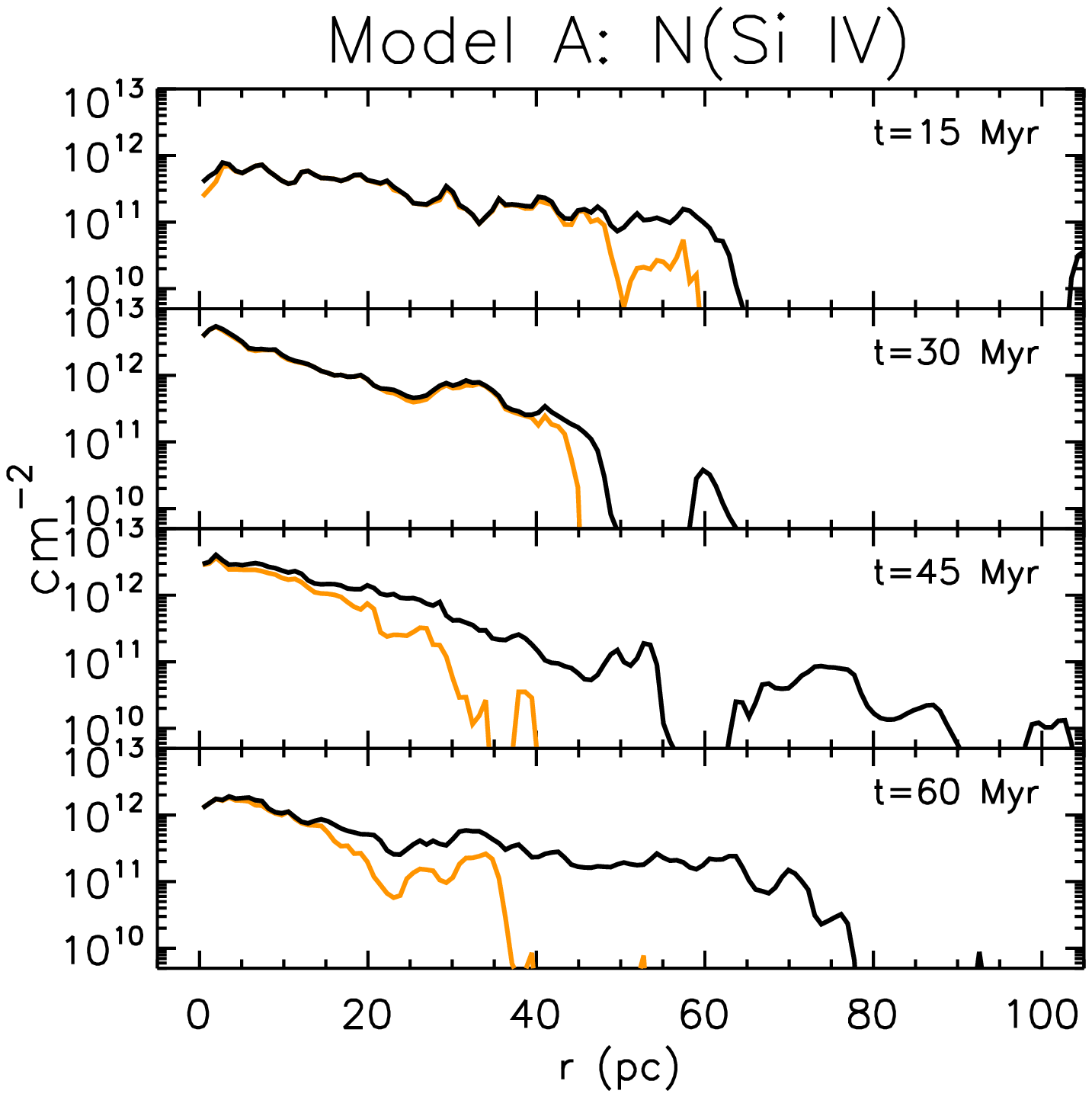}{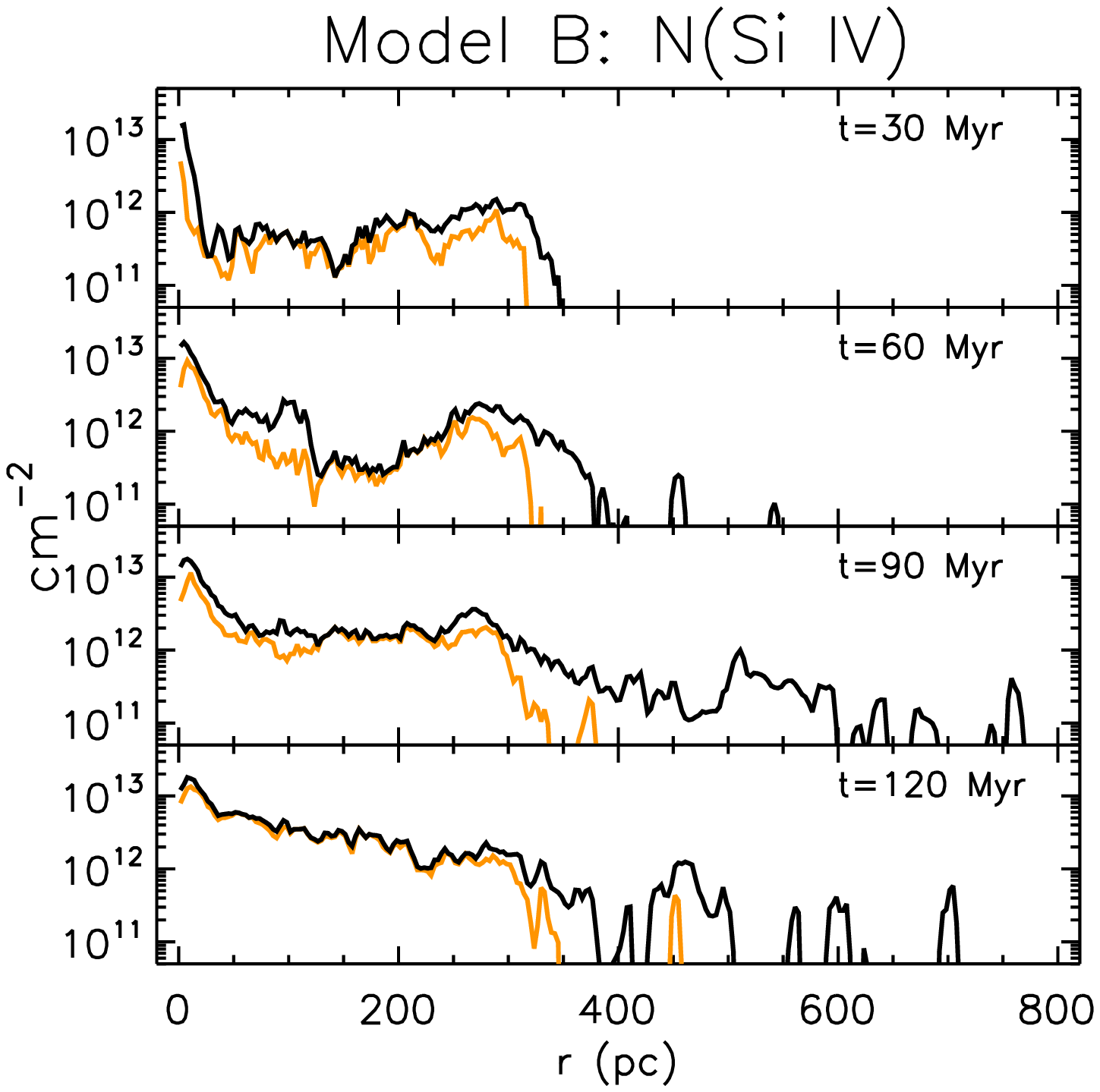}
  \caption{N(Si~IV) as a function of off-axis radius (i.e., distance 
    from the center of the cloud) for HVC Model A (left) and
    HVC Model B (right) at four different epochs (at t= 15, 30, 45, and
    60 Myr for HVC Model A and at t=30, 60, 90, 120 for HVC Model
    B). Note that the initial radii of HVC Model~A and HVC Model~B are 
    approximately 20~pc and 150~pc, respectively, and that 
    t=15 Myr in Model~A corresponds to a similar
    evolutionary phase as t=60 Myr in Model~B. The orange lines track 
    the column densities of the high-velocity material while 
    the black lines track the column densities irrespective of velocity. 
    \label{HVC_NvsR_ModelAB_fig}}
\end{figure}

\begin{figure}
  \epsscale{0.6}
  \centering
  \plotone{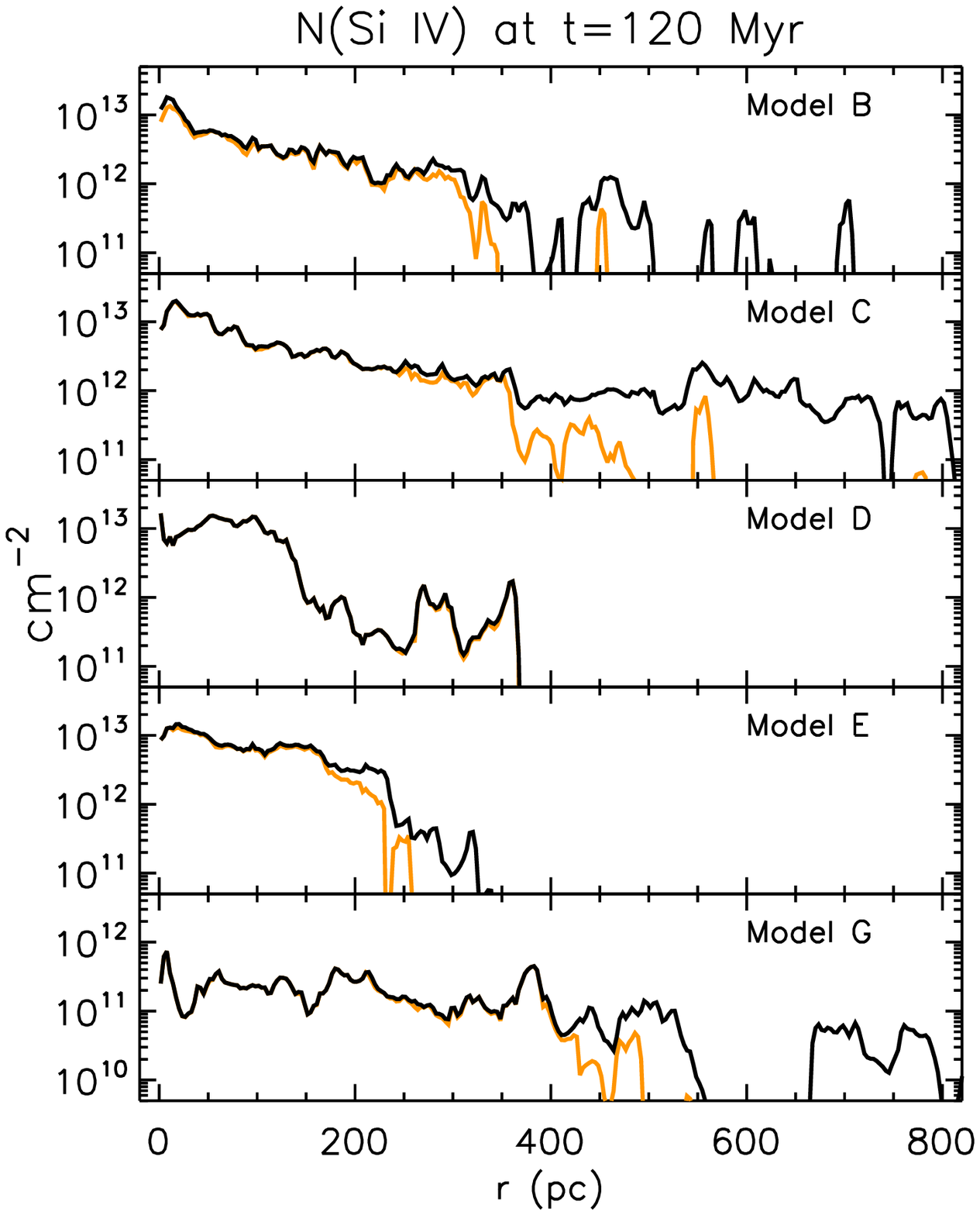} \\
  \plotone{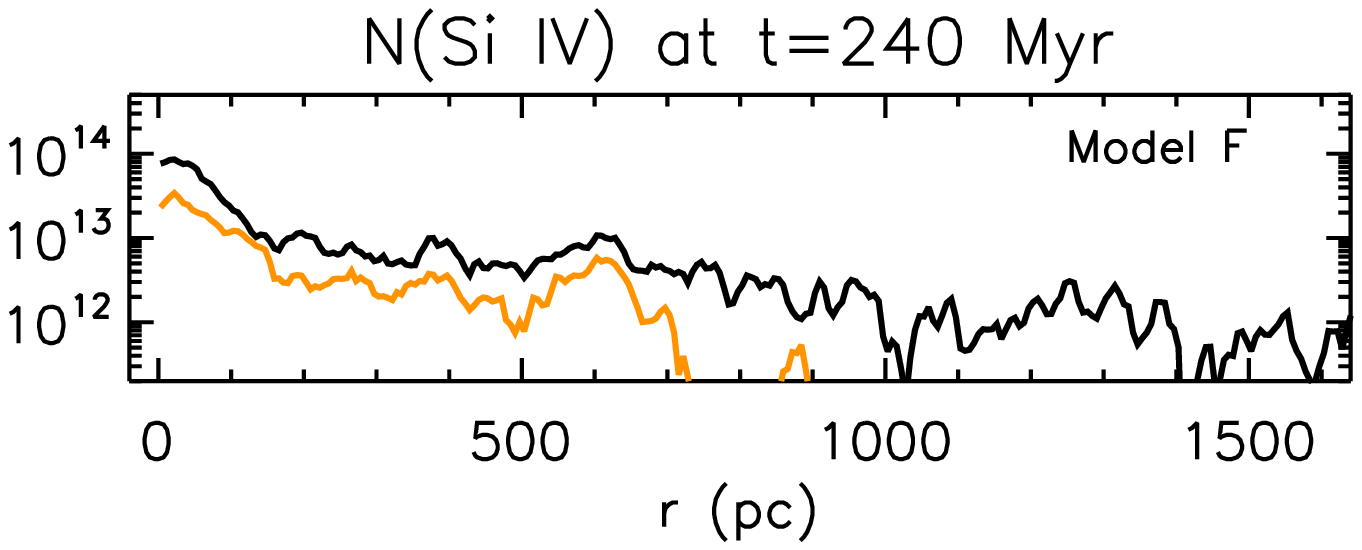}
  \caption{N(Si~IV) as a function of off-axis radius (i.e., distance 
    from the center of the cloud) for our HVC Models (from top
    to bottom, HVC Model B, C, D, E, G, and F) at the last epochs of the
    simulations (i.e., at t=120 Myr for HVC Models B, C, D, E, and G and
    at t=240 Myr for HVC Model F). The orange lines show the 
    column densities of the high-velocity material while 
    the black lines show the total 
    column densities, i.e., the column densities irrespective of velocity.
    \label{HVC_NvsR_others_fig}}
\end{figure}

Next, as in KHS11, we present the 
predicted column densities of high-velocity \ion{Si}{4} 
along vertical sightlines through the simulation domain. 
These column densities can be directly compared with observations. 
We show both the high-velocity 
N(\ion{Si}{4}) and the total, irrespective of velocity, N(\ion{Si}{4}) 
as functions of off-axis radius 
(Figures \ref{HVC_NvsR_ModelAB_fig} and 
\ref{HVC_NvsR_others_fig}: note that we chose the
same epochs for these plots as in Figures $7-9$ in KHS11). 
Figures \ref{HVC_NvsR_ModelAB_fig} and 
\ref{HVC_NvsR_others_fig} show that N(\ion{Si}{4}) is greatest 
along the sightline through the center of the cloud and that it
generally decreases as the off-axis radius increases. A similar trend was 
found in the footprints of N(\ion{C}{4}), N(\ion{N}{5}), and 
N(\ion{O}{6}) (see Figures 7, 8, and 9 of KHS11).  
However, a closer comparison again reveals a greater 
similarity between N(\ion{Si}{4}) and N(\ion{C}{4}), a similarity that 
is likely due to the similarity in the ionization potentials of 
silicon and carbon. For example, in the case of HVC Model~B, 
the footprint and overall shape of the N(\ion{Si}{4}) plot 
(right panel in Figure \ref{HVC_NvsR_ModelAB_fig}) looks more 
similar to that of N(\ion{C}{4}) 
(top right panel in Figure 7 of KHS11) 
than to that of 
N(\ion{O}{6}) (bottom right panel in Figure 7 of KHS11).

\begin{figure}
  \epsscale{1.0}
  \centering
  \plottwo{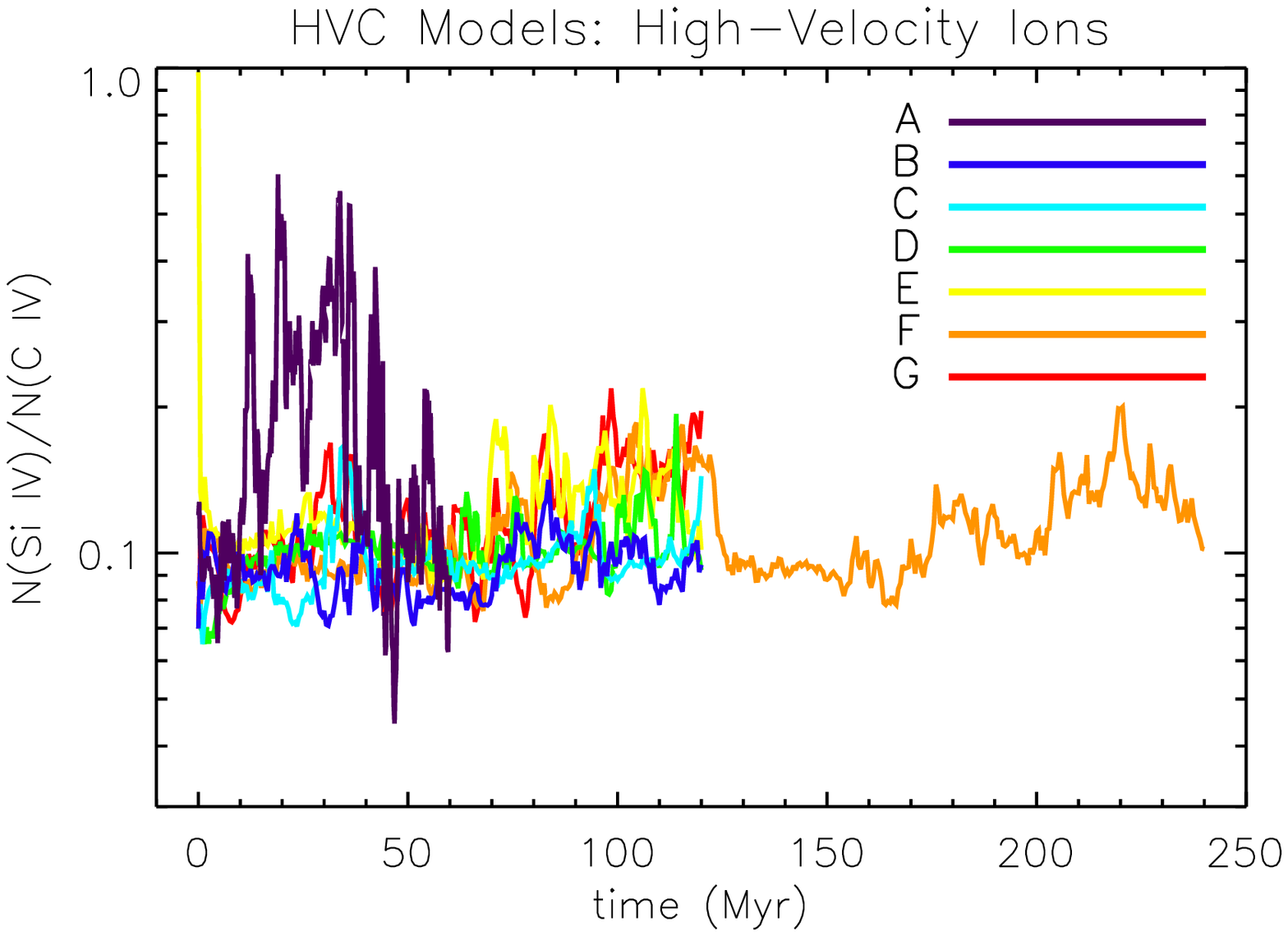}{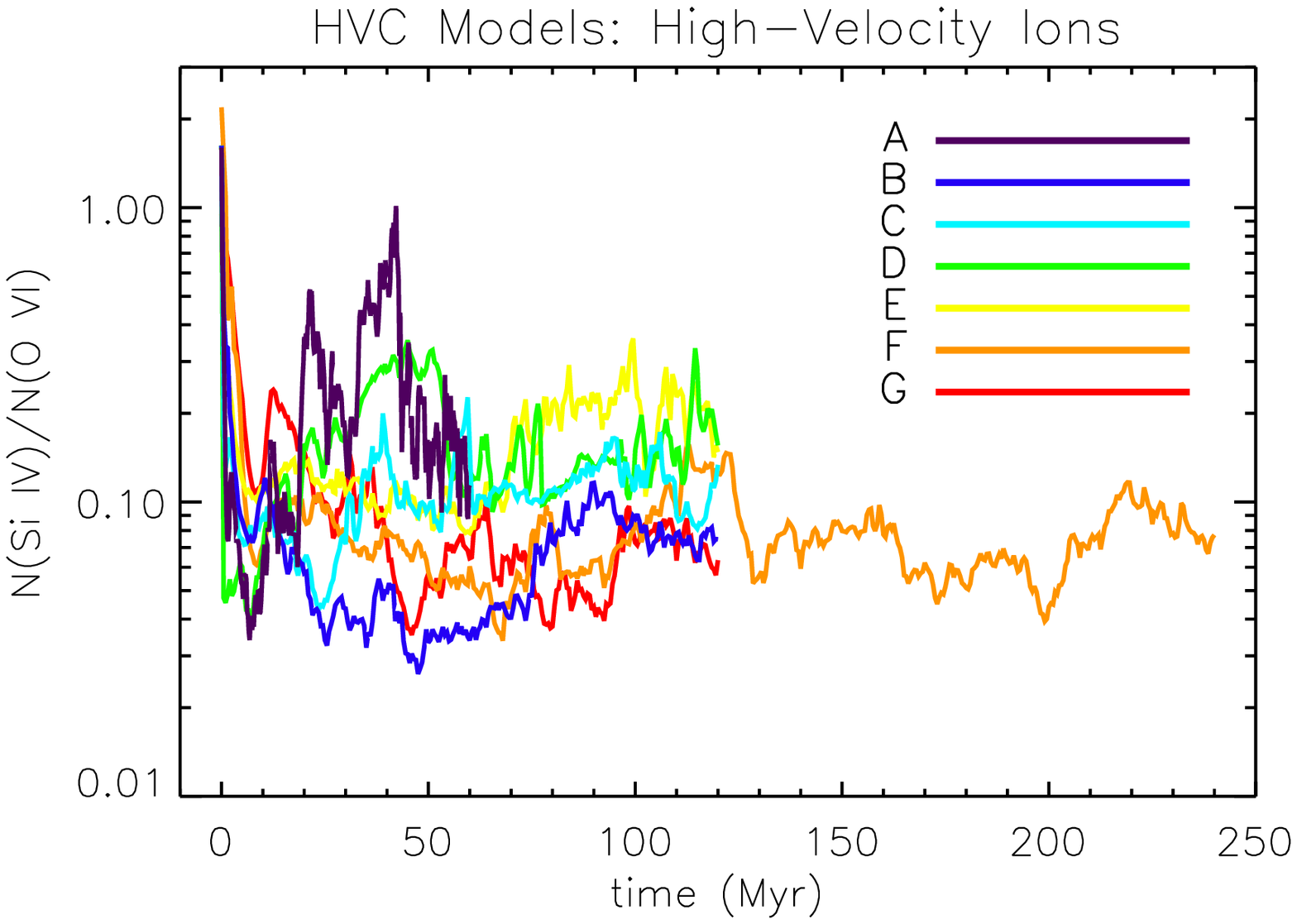}
  \caption{High-velocity N(Si~IV)/N(C~IV) and N(Si~IV)/N(O~VI) 
    as functions of time from the HVC models 
    in the left and right panel, respectively. 
    As for Figure \ref{TML_colm_ratio_fig}, the ratios were 
    calculated from the averages of column densities that were, 
    themselves calculated for many vertical sightlines through 
    the domain. While averaging over these sightlines, we included only 
    the sightlines along which the column density of the relevant 
    high-velocity ion was above a cut-off column density 
    ($10^{11}~\mbox{cm}^{-2}$ for HVC Models B, C, D, E, and F and
    $10^{10}~\mbox{cm}^{-2}$ for HVC Models A and G). 
    \label{HVC_colm_ratio_fig}}
\end{figure}

Not only are the column densities of 
the high-velocity \ion{Si}{4} predicted from 
our HVC models, but also the ratios of the column densities of 
high-velocity \ion{Si}{4} to the column densities of
other high-velocity high ions are of great interest. For this reason, we
present N(\ion{Si}{4})/N(\ion{C}{4}) and
N(\ion{Si}{4})/N(\ion{O}{6}) for the high-velocity ions 
in Figure \ref{HVC_colm_ratio_fig}. 
Note that some material flows out of the domain, but most of this material 
has low velocity (relative to the observer). Thus, the column density 
ratios shown here, like those shown in Figure 11 of KHS11, 
were calculated only from the material within the computational 
domain. We also used the same column density cut-off ($N_{cut}$) 
for the calculation of the average column densities of \ion{Si}{4} 
in order to exclude the contribution from the backgroud gas to the 
average, and the column density ratios are 
taken after the average column densities of individual high-velocity 
high ions are calculated (see equations (2) and (3) in KHS11). 
As can be seen in Figure \ref{HVC_colm_ratio_fig},  
the ratios of high-velocity \ion{Si}{4} to \ion{C}{4} and \ion{O}{6} 
do not vary significantly 
between models, regardless of the variations in model 
parameters in the set of HVC simulations (although HVC Model~A
shows an exceptional trend). 
This also occurred in 
the ratios of high-velocity \ion{C}{4} to \ion{O}{6} and \ion{N}{5} 
to \ion{O}{6} (see Figure 12 in KHS11). 
Table \ref{HVC_colm_ratio_T} shows 
the mean, median, standard deviations, minimum, and maximum ratios obtained
from Figure \ref{HVC_colm_ratio_fig} within a range of time that
excludes the early epochs of each simulation. 

In HKS12, we calculated the quantities of low-velocity \ion{C}{4}, 
\ion{N}{5}, and \ion{O}{6} that the Milky Way halo would acquire 
from infalling HVCs if real HVCs are like those in our simulations. 
These low-velocity high ions are in the gas that ablated from the 
clouds, mixed with the ambient medium, and slowed, sometimes to 
the velocity of the ambient medium. 
The average column density (per species) contribution of low-velocity ions 
due to infalling HVCs
to the observed column densities in the Milky Way's halo
was calculated by using equation (3) of HKS12, 
\begin{equation}\label{eqn1}
\bar{N}(\mathrm{ion}) = \frac{\dot{\mathcal{M}}^{\mathrm{H}}_{\mathrm{HVC}}}{2 \pi 
R^2_{\mathrm{MK}} M^{\mathrm{H~I}}_{\mathrm{HVC,0}} m_{\mathrm{ion}}} 
\int M_{\mathrm{ion}}(t) dt , 
\end{equation}
where $\bar{N}(\mbox{ion})$ is the average column density of a given ion along 
a vertical sightline, 
$\dot{\mathcal{M}}^{\mathrm{H}}_{\mathrm{HVC}}$ 
is the observation-constrained infall rate of high velocity clouds, 
$R_{\mathrm{MK}}$ is the radius of the Milky Way's disk, 
$M^{\mathrm{H~I}}_{\mathrm{HVC,0}}$ is the initial 
\ion{H}{1} mass of each model cloud, 
and $m_{\mathrm{ion}}$ is the atomic mass of the given ion.
We performed the calculation using 
$\dot{\mathcal{M}}^{\mathrm{H}}_{\mathrm{HVC}} = 0.5 M_{\sun}~\mbox{yr}^{-1}$
for consistency with HKS12 and 
$\dot{\mathcal{M}}^{\mathrm{H}}_{\mathrm{HVC}} = 1.0 M_{\sun}~\mbox{yr}^{-1}$
in light of more recent infall rate estimates (see Section \ref{compare_S_lowvel}).
We took $R_{\mathrm{MK}}$ to be $25$~kpc as in HKS12.
Following the same approach that HKS12 used for 
\ion{C}{4}, \ion{N}{5}, and \ion{O}{6} (see Section 4 of HKS12), 
we first integrated $M_{\mathrm{Si~IV}}(t)$ (shown in Figure \ref{si4_mass_evol_fig}) 
with respect to time over the whole length of each simulation. Then, 
we used the resulting integral in Equation (\ref{eqn1}) which 
estimates the average column density 
of low-velocity \ion{Si}{4} in the Milky Way's halo.

The use of hydrocode simulation results in this calculation 
is complicated by the facts that
some of the mixed gas flows out of the domain late 
in the simulated timespan and that the
simulation timespan ends before the high ions have completely disappeared. 
We have attempted to compensate for these effects 
when preparing our tabulations of the average column 
densities (Table \ref{low_vel_si4_halo_T}).

We present four families of estimates. 
The first two estimates are 'Domain only' 
(column [2] in Table \ref{low_vel_si4_halo_T}) and 
`Domain + Escaped' (column [3] in Table \ref{low_vel_si4_halo_T}) 
under the category 'Average N(\ion{Si}{4}) during the simulations'.
As in HKS12, the first estimates, i.e., 
the column densities listed under the category 
'Average N(\ion{Si}{4}) during the simulations' and 
labeled `Domain only' 
(column [2] in Table \ref{low_vel_si4_halo_T}) use the 
low-velocity \ion{Si}{4} only present in the simulational domain 
(i.e., the time integration is calculated from the 
blue dotted lines in Figure \ref{si4_mass_evol_fig}), 
while the second estimates, i.e.,
the column densities labeled `Domain + Escaped' 
(column [3] in Table \ref{low_vel_si4_halo_T}) 
use this `Domain only' \ion{Si}{4} plus low-velocity \ion{Si}{4} that 
has escaped from the domain during the simulation time 
(i.e., the time integration is 
calculated from the blue solid lines in Figure 
\ref{si4_mass_evol_fig} during the simulation time). 
Because 'Domain + Escaped' ion column 
densities take into account the 'Escaped' ions, 
the `Domain only' and `Domain+Escaped' ion column densities 
generally correspond to lower and upper limits, respectively, 
of the contribution from the clouds during the simulation time.

However, the first two estimates, i.e., the column densities 
listed under the category 'Average N(\ion{Si}{4}) during the simulations',
are still smaller than 
the true column densities because the simulations ended 
before all the material of the initial cloud 
(assumed to be neutral, initially) disappeared. 
Therefore, in addition to these quantities, we also present the last two 
columns of estimates of the four families, i.e., the average \ion{Si}{4} column densities 
calculated with the cloud's lifetime taking into account: 'Domain only' 
(column [5] in Table \ref{low_vel_si4_halo_T}) and `Domain + Escaped' 
(column [6] in Table \ref{low_vel_si4_halo_T}) under the category 
'Average N(\ion{Si}{4}) during the cloud's lifetime'. 
In order to account for the clouds' contributions throughout 
their entire lives, we followed HKS12 and 
assumed that the amounts of low-velocity 
high ions produced by the cloud are proportional to the mass of the cloud 
that is ``removed'' from the initial cloud by either ablation or ionization. 
In this case, the time integral $\int M_{\mathrm{Si~IV}}(t)dt$ (integrated 
over the entire life of the cloud) in Equation (\ref{eqn1}) can be replaced 
by $\int_0^{T_{\mathrm{sim}}}M_{\mathrm{Si~IV}}dt/\beta_{\mathrm{HVC}}$, where 
$\beta_{\mathrm{HVC}}=M_{\mathrm{HVC,lost}}^{\mathrm{H~I}}/M_{\mathrm{HVC,0}}^{\mathrm{H~I}}$ is 
the fraction of the initial cloud mass that is ``lost'' 
due to either ablation or ionization by the end of the simulation (HKS12). 
Thus, the estimated column densities
(columns [2] and [3] in Table \ref{low_vel_si4_halo_T}) 
divided by $\beta_{\mathrm{HVC}}$ (tabulated in both Table 2 of 
HKS12 and Table \ref{low_vel_si4_halo_T} of this paper) yield 
better estimates. Columns [5] and [6] in Table \ref{low_vel_si4_halo_T} are 
obtained by dividing columns [2] and [3] by  $\beta_{\mathrm{HVC}}$, 
respectively.
These values for \ion{Si}{4} can be compared with the results for 
\ion{C}{4}, \ion{N}{5}, and \ion{O}{6} in Table 2 of HKS12.

\section{Comparison with Observations and Other Models}
\label{compare_S}

In this section, we compare our predicted \ion{Si}{4}
column densities with observations and predictions from other
models. Gas in and near the
Galaxy is generally classified as high velocity 
($|v_{LSR}| \gtrsim 90~\mbox{km}~\mbox{s}^{-1}$) or
low velocity ($|v_{LSR}| \lesssim 30~\mbox{km}~\mbox{s}^{-1}$). 
(Although velocities between 30 and 90~km~s$^{-1}$ are also commonly termed intermediate
velocities, for convenience, we include this range in our low velocity range.) 
Here, we make the same distinction, comparing our results with other models 
and observations of high velocity ions in Section \ref{compare_S_highvel} 
and comparing our results with other models and observations
of low velocity ions in Section  \ref{compare_S_lowvel}. 
While making our predictions, we consider only the
vertical component of the velocity along vertical sight lines through the simulational domains.
Only the HVC simulations have \ion{Si}{4} with large vertical velocities. 
Both the HVC and the TML simulations have \ion{Si}{4} with low vertical velocities, 
because material ablated from the cloud 
is able to decelerate in the HVC simulation and because the cloud's 
motion is in the horizontal direction in the TML simulation.


\subsection{High-Velocity Si~IV}
\label{compare_S_highvel}

For the observed column densities of high-velocity \ion{Si}{4}, we
use the results of \citet{Shulletal2009ApJ}, where column densities
of high- and intermediate-velocity \ion{Si}{4} 
(along with \ion{Si}{2} and \ion{Si}{3}) toward 37 AGNs at 
high Galactic latitudes were reported. Note that the study 
of \citet{Shulletal2009ApJ} is the largest collection of 
high-velocity \ion{Si}{4} measurements to date.
For comparison with the predictions of our HVC models, 
we consider only the detected high-velocity components in Table 2
of \citet{Shulletal2009ApJ}. 
There are 37 such detections (note that on some sightlines
there are no detections, and on others more than one). 
The average, median, standard deviation, 
minimum, and maximum of logarithmic column densities of 
high-velocity \ion{Si}{4}, log[N(\ion{Si}{4})], 
are 12.81, 12.77, 0.34, 11.98, and 13.51, respectively, for 
the sample composed of 37 detections (i.e., excluding non-detections 
and counting mutiple detections separately along the same sightline). 

Comparing such values with our simulational results is not so simple. 
For a collection of vertical sightlines through the simulational domain, 
in general, the predicted high-velocity N(\ion{Si}{4}) 
peaks near the central sightline, 
remains sizeable for sightlines within two times the initial cloud radius, 
and decreases along the sightlines at larger off-axis radii 
from the central sightline 
(Figures \ref{HVC_NvsR_ModelAB_fig} and \ref{HVC_NvsR_others_fig}). 
Although the predicted column densities of high-velocity \ion{Si}{4} 
vary along wide ranges of sightlines at different epochs, some of them 
are consistent with the observations. 
For example, in cases of HVC Models~B, C, D, and E 
(whose initial cloud radii are $\sim150$~pc), 
the predicted high-velocity N(\ion{Si}{4}) reaches slightly above 
$10^{13}~\mbox{cm}^{-2}$ along the central sightline (i.e., $r\sim0$~pc), 
especially at later times. 
Along the sightlines at larger off-axis radii in these models 
(i.e., $r\sim150-200$~pc) at $t=120$~Myr, the predicted high-velocity
N(\ion{Si}{4}) is still above $10^{12}~\mbox{cm}^{-2}$, which is also 
consistent with the minimum observed value. Similar overlap between 
predictions for sightlines that pass vertically 
through a single HVC’s shroud 
and observations can be found for HVC Model~A 
(whose initial cloud radius is $\sim20$~pc), HVC Model~G 
(which has an order of magnitude lower density than HVC 
Models~B to E), and HVC Model~F (which has the largest radius, $\sim300$~pc), 
for a range of sightlines at some selected times.

However, note that the central sightline is only a small fraction of 
the cross section of the modeled HVC structure and that 
the column densities along sightlines at larger off-axis radii from the central 
sightline are often outside the observed range. Thus, if we were to take into 
account the effect of averaging the predicted column densities over all of the 
sightlines that intersect the cloud 
(instead of comparing the predicted column density along a specific 
sightline at a given time with the observations as above), 
reality would often show higher column densities than our HVC models.
This is shown in Figure \ref{hvc_si4_comp_obs_fig}, 
where we compare the observed distribution of \ion{Si}{4} column densities 
with the average \ion{Si}{4} column
density of high vertical velocity gas in our HVC models as a function of time. 
To distinguish sightlines that contain \ion{Si}{4} from those that do not, 
we excluded sightlines whose \ion{Si}{4} column
densities fall below our cut-off ($N_{cut}$), which is $10^{11}~\mbox{cm}^{-2}$ 
for HVC Models B through F and $10^{10}~\mbox{cm}^{-2}$ for HVC Models A and G.

Even regions in our modeled HVC structures having only marginal 
amounts of \ion{Si}{4} could contribute to the observed 
\ion{Si}{4} column densities if multiple HVC structures exist 
along the observed lines of sight. Such HVCs need not be on the 
size scale of the named complexes, but could be cloudlets within 
a complex, and the structures could include ablated material behind
the infalling clouds. The ablated material may be ionized and so 
not readily counted in \ion{H}{1} surveys. Although it is possible 
to postulate such a multiple-cloud model, the number of HVC 
structures along any given line of sight could be difficult to 
constrain. For this reason, it is more practical to compare the 
ratios of the \ion{Si}{4}, \ion{C}{4}, and \ion{O}{6} column 
densities to those garnered from the observations and those 
extracted from other models.

Figure \ref{si4o6_c4o6_high_fig} compares the observed 
N(\ion{Si}{4})/N(\ion{O}{6}) and N(\ion{C}{4})/N(\ion{O}{6}) ratios 
with those from our models and from other models, 
a shock heated gas model and two radiatively cooling gas models. 
First of all, our HVC models predict the observations very well. 
Some predictions from other models, such as 
a shock heating model 
\citep[][two data points in the upper-right region 
correspond to their models having the highest two 
shock speeds, 400 and 500 km/s, and largest magnetic field]
{DopitaSutherland1996ApJS} are close to the 
observations, too. 
Note that although the radiative cooling model including 
the isochoric non-equilibrium ionization 
\citep[][dashed orange line]{GnatSternberg2007ApJS} matches 
the observed ratios of high-velocity ions 
when the temperature reaches an approximate range of 
$(1.4 - 2.2) \times 10^4$ K 
(note that the temperature decreases along all three orange lines 
from lower-left to upper right), these predictions are more 
relevant to the ratios of low-velocity ions which we will discuss 
in the following section.

\subsection{Low-Velocity Si~IV}
\label{compare_S_lowvel}

Both our TML and HVC models predict low-velocity \ion{Si}{4} that
can be compared with observed \ion{Si}{4} in the halo of the Milky Way. 
In our TML simulations,
in order to sample the mixed zone most effectively, our simulated lines of sight 
pass perpendicularly through the mixed zone and are perpendicular to the direction 
of motion of the cool cloud. 
As a result, all of the \ion{Si}{4} in the domain appears to have low velocities.
In contrast, the low-velocity \ion{Si}{4} in our HVC simulations,
resides in mixed gas that has slowed to low-velocities ($< 80~\mbox{km}~\mbox{s}^{-1}$ 
for HVC Models A, B, F, and G and $< 100~\mbox{km}~\mbox{s}^{-1}$ for HVC Models 
C, D, and E)
and is observed along sightlines that are parallel to the direction of motion of the
HVC.  We first compare the low-velocity \ion{Si}{4} observed on Milky Way halo lines
of sight with those predicted from our TML simulations and then we compare with
the column density of decelerated \ion{Si}{4} expected to result 
from a population of HVCs.
The ratios of \ion{Si}{4} to \ion{O}{6}, \ion{Si}{4} to \ion{C}{4}, 
and \ion{C}{4} to \ion{O}{6} column densities
form the basis of another important test, which we then perform. 
Furthermore, the high ions in the halo may be due to multiple causes and 
therefore, later in this section, we add our new \ion{Si}{4} results 
to the composite scenario for low-velocity
high ions in the halo that was presented in HKS12 and compare with observations.

To determine the average column density of 
low-velocity \ion{Si}{4} observed in the Milky Way, 
we use the results of \citet{Wakkeretal2012ApJ}, where
column densities of low-velocity \ion{Si}{4} (along with \ion{O}{6}, \ion{N}{5},
\ion{C}{4}, and \ion{Fe}{3}) toward 58 extragalactic
objects were reported. Note that the study of \citet{Wakkeretal2012ApJ} includes 
the largest number of observed sightlines for the low-velocity high ions to date. 
Along a given observed line of sight, 
\citet{Wakkeretal2012ApJ} presented both the component column densities within 
certain ranges of velocities and the integrated column densities which sum up 
the component column densities within a wider range of velocities. Because 
the predicted column densities from our model calculations are also integrated 
over a wide range of velocities, we compare our predictions with the integrated 
column densities from \citet{Wakkeretal2012ApJ}. 
The average of the integrated \ion{Si}{4} column
densities from \citet[][not latitude corrected]{Wakkeretal2012ApJ} 
is $\sim 3.7 \times 10^{13}~\mbox{cm}^{-2}$, while the typical 
\ion{Si}{4} column density predicted from sightlines oriented
perpendicular to the single hot-cool interface in each of our 
TML Models~A, B, D, E, and F 
is about $10^{12}~\mbox{cm}^{-2}$. 
In TML Model C (which examined turbulent mixing on smaller scales over 
shorter time periods and has an order of magnitude 
smaller domain than the other TML models), 
the average \ion{Si}{4} colum density calculated 
over the time interval of $[6,8]$~Myr is $\sim~1.3 \times 10^{11}~\mbox{cm}^{-2}$. 
(Note that at a given time, the spatially averaged column density 
is calculated first and then the temporal average of spatially averaged 
column densities over the given time intervals is calculated. The average 
column density for TML Model~C is calculated in this way.) 
If all of the observed low-velocity \ion{Si}{4} in the halo is due to
TMLs, then 37 layers like those in TML Models A, B, D,
E, and F or 280 layers like those in TML Model C, 
are required to be along the average line of sight\footnote{
The average integrated column densities of \ion{C}{4} and \ion{O}{6} 
obtained from 
\citet{Wakkeretal2012ApJ} are $\sim 1.5 \times 10^{14}~\mbox{cm}^{-2}$ 
and $\sim 2.0 \times 10^{14}~\mbox{cm}^{-2}$, respectively, while 
the predicted column densities of these two ions from our TML models 
(except TML Model~C) 
are $\sim 7.3 \times 10^{12}~\mbox{cm}^{-2}$ 
and $\sim 4.4 \times 10^{12}~\mbox{cm}^{-2}$, respectively. 
Thus, approximately 20 and 45 layers are required to explain the observed 
column densities of the \ion{C}{4} and \ion{O}{6} ions, respectively.
}. 
Note that, because the low-velocity \ion{Si}{4} scale height is a few
kiloparsecs \citep{SavageWakker2009ApJ}, and the mixing layers in our TML models
are typically $\sim100$ pc thick ($\sim10$ pc for TML Model C), it may be
difficult to fit the required number of mixing layers along the line of
sight.

We now consider the column densities of 
low-velocity \ion{Si}{4} in decelerated gas ablated from HVCs. 
Here, rather than examining the contribution of a single HVC, 
we consider the effect of an ensemble  
of HVCs that are like our model HVCs
(i.e., the model calculated in Section \ref{HVC_si4_S} and tabulated in 
Table \ref{low_vel_si4_halo_T}).
The predicted column densities are compared with the average observed column 
density in Figure \ref{si4_halo_fig}. 
For the average observed column density, 
we again use the \ion{Si}{4} results from \citet{Wakkeretal2012ApJ}. 
Figure \ref{si4_halo_fig}
shows that the predicted low-velocity N(\ion{Si}{4}) severely 
underpredicts the observed N(\ion{Si}{4}). Even the largest 
prediction ($\sim 5.4 \times 10^{12}~\mbox{cm}^{-2}$ 
estimated with an HVC infall rate 
$\dot{\mathcal{M}}^{\mathrm{H}}_{\mathrm{HVC}}=1.0 M_{\sun}~\mbox{yr}^{-1}$) 
which is from a population of clouds like those in 
HVC Model F and includes both the ions that have escaped 
from the domain and the effect of the lifetime evolution 
of the cloud, is smaller than the average observed value
by a factor of $\sim 5$. 
A similar comparison for the \ion{C}{4}, \ion{N}{5}, and \ion{O}{6} ions
was done in HKS12 (with an 
HVC infall rate 
$\dot{\mathcal{M}}^{\mathrm{H}}_{\mathrm{HVC}}=0.5 M_{\sun}~\mbox{yr}^{-1}$), 
where we concluded that the model was better at explaining the observed 
\ion{O}{6} than at explaining the 
observed \ion{N}{5} or \ion{C}{4} 
(this conclusion does not change even with a factor 2 larger infall 
rate $\dot{\mathcal{M}}^{\mathrm{H}}_{\mathrm{HVC}}=1.0 M_{\sun}~\mbox{yr}^{-1}$).

As mentioned earlier, comparing the ratios of column densities of multiple 
high ions, such as \ion{Si}{4} to \ion{O}{6}, 
between observations and predictions is a better method for 
constraining the models. 
The comparison for the low-velocity high ions is shown in 
Figure \ref{si4o6_c4o6_low_fig}, including the predictions from our TML models 
and the decelerated gas in our HVC simulations. First of all, it 
is interesting to recognize that the observed ratios are clustered 
more tightly than the model predictions. 
In this plot, the predictions from 
our TML models (purple triangles) and the galactic fountain model 
(cyan squares) are in good agreement with the observations. However, 
careful comparison between the predictions of our TML models and 
the observations shows that the predicted N(\ion{C}{4})/N(\ion{O}{6}) 
ratio is slightly larger than the central value of the observed ratios by about 
0.3 dex, while the predicted N(\ion{Si}{4})/N(\ion{O}{6}) ratio is similar to 
the central value of the observed ratios. Note that this is consistent with 
the disparity in the aforementioned estimates of the numbers of layers that are 
required to explain the observed column densities of these three ions, i.e., 
20, 37, and 45 layers for \ion{C}{4}, \ion{Si}{4}, and \ion{O}{6}, respectively.
Some predictions from supernova remnant models \citep[][green 
triangles]{Shelton2006ApJ} at late phases of the evolution overlap or 
are close to the observations, but there is a tendency for this model to 
slightly underpredict the N(\ion{Si}{4})/N(\ion{O}{6}) ratio. 
The predictions from the isochoric radiative cooling model that includes non-equilibrium 
ionization \citep[][dashed orange line]{GnatSternberg2007ApJS} match 
the observations well in the $(1.4 - 2.2) \times 10^4$ K temperature range 
(note again that the temperature 
decreases along all three orange lines from lower-left to upper right).  
All the other models including our HVC models (blue squares) 
either significantly underpredict the N(\ion{Si}{4})/N(\ion{O}{6}) ratio
or, as in the cases of the CIE radiative cooling 
\citep[][at $\mbox{T} \gtrsim 1.7 \times 10^5 ~\mbox{K}$]
{GnatSternberg2007ApJS,SutherlandDopita1993ApJS}, 
overpredict the N(\ion{C}{4})/N(\ion{O}{6}) ratio.

Recently, \citet[][their Section 5, in particular, their Figure 11]
{Wakkeretal2012ApJ} made a detailed comparison 
for the ratios of column densities of low-velocity high ions 
between various model predictions 
and observation data sets that they collected and analyzed. 
What they found from their comparison is quite similar to our results
discussed above\footnote{
\citet{Wakkeretal2012ApJ} re-scaled all the model 
predictions according to the abundances of \citet{Asplundetal2009ARAA}
which have more carbon and silicon by 0.05 and 0.24 dex, respectively, 
than those of \citet{Wilmsetal2000ApJ}. The abundances of oxygen are 
almost identical in both references.}: 
our TML models, the galactic fountain, 
and the non-equilibrium radiative cooling 
model (with $T\sim 1-2 \times 10^4$~K) are favored. 
Note that in all three of these models 
ionization of interesting atoms is calculated 
in the non-equilibrium fashion (this is also true for the 
case of the ratios of high-velocity high ions). 
Thus, including the non-equilibrium ionization calculation is essential 
in the modeling of whatever physical processes are 
relevant to the production of high ions.

As shown in Figure \ref{si4o6_c4o6_low_fig}, our models of the 
decelerated gas following HVCs underpredict 
the column densities of low-velocity \ion{C}{4} and \ion{Si}{4} with respect to 
those of low-velocity \ion{O}{6}. Note that both \ion{C}{3} and \ion{Si}{3} 
ions are more susceptible to photoionization than the \ion{O}{5} ions, but 
our HVC simulations have the limitation that they did not include 
the effect of the photoionization. 
Photoionization is astrophysically important and suspected of boosting 
the \ion{C}{4} and \ion{Si}{4} contents of the gas over those predicted
in our simulations. 
For this reason, we suggested a composite model in HKS12 
which combines together four model sources for 
low-velocity high ions, in order to attempt to 
explain the full amounts of the observed low-velocity high ions.
Note that in HKS12 we did not expect that our HVC models alone 
could explain the full amounts of observed low-velocity high ions. 
(We would like to remind readers that the primary 
purpose of our estimating the amounts of low-velocity high-ions from 
our HVC models in HKS12 is to quantitatively measure the contribution 
from this model to the total amounts.) 
The composite model was composed of contributions from the following 
four sources of high ions: slowed ablated gas from a population of HVCs 
like our HVC Model~B, a population of 
extraplanar supernova remnants \citep{Shelton2006ApJ}, 
galactic fountains \citep{ShapiroBenjamin1993sfgi.conf}, 
and photoionization by an external 
radiation field \citep{ItoIkeuchi1988PASJ}. Choosing 
these four sources may be justified by a dynamic picture of the Galactic 
halo, where these four sources work as independent sources of high ions 
while disturbing the halo in different ways. 

Table \ref{composite_T} presents the updated results of 
our composite model, with the new \ion{Si}{4} calculations included 
(cf. Table 3 of HKS12). First of all, note that the composite model 
is designed to reproduce the observed \ion{O}{6} column density. 
In order to reflect the most updated observational 
data, we used the results from \citet{Wakkeretal2012ApJ}. 
We calculated the average $N \sin |b|$ for all four ions from their Table 2, 
considering only the integrated column densities along the detected sightlines. 
Note also that all of the model predictions in Table \ref{composite_T} 
are re-scaled using the elemental abundances of \citet{Wilmsetal2000ApJ} 
(in Table 3 of HKS12, the re-scaling has not been done to the model predictions).
However, these adjustments do not alter the conclusions of HKS12: 
the composite model reproduces most of the observed \ion{C}{4} 
column density, but underpredicts the observed \ion{N}{5} column 
density. The model also underpredicts the observed \ion{Si}{4} 
column density; the discrepancy between the predicted and observed 
values is more severe than for \ion{N}{5}. 
(The contribution to \ion{Si}{4} from the SNR model 
was not included in Table \ref{composite_T} because it was not available. 
However, its contribution is not expected to be large enough to 
change the overall conclusion.) 
The severe shortage of \ion{Si}{4} in the composite model 
occurs because the models that exclude photoionization 
tend to underpredict \ion{Si}{4} to an extent that is not fully 
compensated for by the chosen photoionization model. 
However, the referenced photoionization model 
\citep[i.e.,][]{ItoIkeuchi1988PASJ}, was not well constrained, 
allowing the possibility that more \ion{Si}{4} has resulted from 
photoionization than thought.

The amounts of high ions produced via photoionization depend on the 
spectrum of ionizing photons. For example, \citet{ItoIkeuchi1988PASJ} 
used a simple power-law spectrum such as $I(\nu) = I_0 (\nu / \nu_T)^{-1}$, 
where $I_0$ is the intensity at the Lyman limit of hydrogen, $\nu_T$. 
They also assumed that this spectrum does not change spatially in the Galactic 
halo. However, it may be possible to increase the amount of \ion{Si}{4}, only, 
without affecting those of the other high ions (especially, the amount of 
\ion{C}{4} which is also susceptible to photoionization) by changing the 
spectrum of ionizing photons. Note that 
the ionization potential of \ion{Si}{3} (33 eV) is slightly 
lower than that of \ion{C}{3} (48 eV). 
Thus, if the spectrum of ionizing photons 
were dominated by photons with energy between these two ionization potentials, 
only the amount of \ion{Si}{4} would increase. 
Testing whether such a spectrum is physically realistic, and whether
such a spectrum could indeed bring the \ion{Si}{4} column density in line with
the observations, is beyond the scope of the current paper, but could be
addressed in a future study.

Finally, it is worth mentioning again the effect of uncertain infall rates of
HVCs on the predictions of low-velocity high ions that were made from our
HVC models. HKS12 quoted several observationally based estimates.
In addition to those, we would like to add 
$\sim 0.1 ~ M_{\sun}~\mbox{yr}^{-1}$ from \ion{H}{1} HVCs \citep{PutmanPeekJoung2012ARAA}, 
$0.1 - 1.4 ~ M_{\sun}~\mbox{yr}^{-1}$ from ionized HVCs \citep{LehnerHowk2011Sci}, and
$\sim 1 ~ M_{\sun}~\mbox{yr}^{-1}$ from \ion{Si}{3} HVCs \citep{Shulletal2009ApJ}.
Furthermore, a recent estimate of the ionized 
mass of the Magellanic System implies that the mass infall rate of this system 
alone is as large as $\sim 3.7 - 6.7 ~ M_{\sun}~\mbox{yr}^{-1}$ \citep{Foxetal2014ApJ}. 
In the current paper, we use both 
$\dot{\mathcal{M}}^{\mathrm{H}}_{\mathrm{HVC}}=0.5 M_{\sun}~\mbox{yr}^{-1}$, 
for consistency with our earlier work in HKS12 and 
$\dot{\mathcal{M}}^{\mathrm{H}}_{\mathrm{HVC}}=1.0 M_{\sun}~\mbox{yr}^{-1}$, 
in light of the more recent estimates.

The effect of increasing the mass infall rate is to increase the column densities
of infalling high ions by a factor of 2, such that the values in 
Figure \ref{si4_halo_fig} and Table \ref{low_vel_si4_halo_T} 
are doubled, while the ion ratios for infalling material are not affected. 
The effect on the composite model is slightly complicated because the contribution 
from galactic fountains is calculated based on the assumption that the 
observationally determined average \ion{O}{6} column density would equal the sum
of the contributions from HVCs, extraplanar SNRs, and 
galactic fountains. Therefore, the contribution from galactic fountains decreases 
when the assumed HVC infall rate increases.
The contributions from extraplanar SNRs and photoionization
by an external radiation field are not affected by the change in assumed HVC
infall rate and photionization does not contribute to the \ion{N}{5} 
or \ion{O}{6} column densities. 
In Table \ref{composite_T},
we present the various components of our composite model for both
the originally assumed HVC infall rate and the revised, higher rate (these new
predictions are in parentheses underneath the original predictions).
By affecting the calculated fountain contribution, the increased infall rate decreases 
the total \ion{Si}{4} and \ion{C}{4} column
densities predicted by the composite model, while increasing the total
predicted \ion{N}{5} column density. Thus, the revised composite model does not 
account for the observed \ion{C}{4} as well as the original.

\section{Summary}\label{summary_S}

We have presented the column densities of \ion{Si}{4} predicted 
from two sets of hydrodynamic simulations: TML and HVC simulations. 
In the TML simulations which are carried out 
in 2D Cartesian coordinates, 
the column densities are calculated 
along the lines of sight that are perpendicular to the initial 
interface between the cool cloud and the hot ambient gas. 
The HVC simulations, carried out in 2D cylindrical coordinates, 
model cool spherical clouds falling through a hot ambient medium. 
In the HVC simulations, the column densities 
are calculated along the vertical lines of sight parallel to the 
initial velocity vector of the cloud assuming that the observers 
are located below the clouds. 
As in KS10, KHS11, and HKS12 which presented the \ion{C}{4}, 
\ion{N}{5}, and \ion{O}{6} predictions, the quantities of \ion{Si}{4} ions 
in our simulations were calculated in a non-equilibrium fashion, 
i.e., the ionization states were updated at each 
hydrodynamic time step. 
Our calculations did not include photoionization. As a result, 
the predicted quantities of \ion{Si}{4} can be seen as lower 
bounds on the quantities that would exist in an environment that is 
affected by both the hydrodynamics of mixing and photoionization.

In both the TML and the HVC simulations, 
\ion{Si}{4} became 
abundant in the mixed gas that forms via mixing of cool cloud gas 
with the hot ambient gas, just as \ion{C}{4}, \ion{N}{5}, and 
\ion{O}{6} did. 
The general features of the predicted 
\ion{Si}{4} ions in both the TML and the HVC simulations are more 
similar to those of \ion{C}{4} than to those of \ion{N}{5}
and \ion{O}{6}. There are three examples of this:
(1) the time evolution of the high ion column densities in the TML 
models (Section \ref{TML_si4_S}), 
(2) the growth in quantities of high ions as a function of time 
in the HVC models (Section \ref{HVC_si4_S}), 
and (3) the ion column densities as a 
function of off-axis radius in the HVC models (Section \ref{HVC_si4_S}). 
This is likely because neutral silicon ($1s^2 2s^2 2p^6 3s^2 3p^2$) 
has a similar electron configuration to neutral carbon ($1s^2 2s^2 2p^2$)
thus the ionization process to ionize \ion{Si}{3} to \ion{Si}{4} 
is more similar to that to ionize \ion{C}{3} to \ion{C}{4} 
than to those for \ion{N}{4} to \ion{N}{5} or \ion{O}{5} to \ion{O}{6}.

\begin{figure}
\centering
\epsscale{0.7}
\plotone{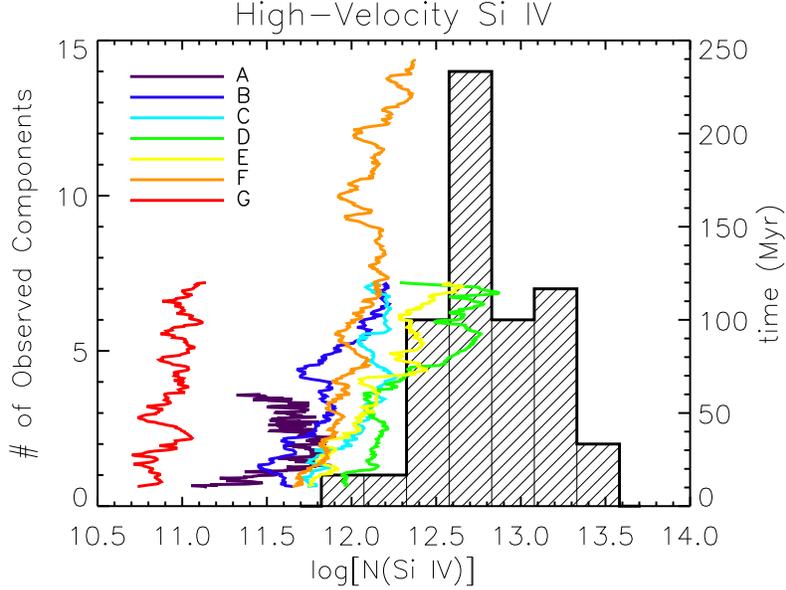}
\caption{
  Column densities of high-velocity \ion{Si}{4}: comparison between 
  observations and predictions from our HVC models. The gray histogram 
  presents 37 detected components of high-velocity \ion{Si}{4} 
  column densities obtained from Table 2 of \citet{Shulletal2009ApJ}. The 
  number of observed components (in the left y-axis) within a range of 
  logarithmic column densities of high-velocity \ion{Si}{4} (in the 
  x-axis) is plotted. The time evolution (from $t=10$ Myr) 
  of average column densities 
  of high-velocity \ion{Si}{4} predicted from our seven HVC models is 
  shown as solid color lines (the right y-axis versus x-axis). The average 
  column density of high-velocity \ion{Si}{4} at each epoch is 
  calculated according to equations 
  (2) and (3) in KHS11 with the same cut-off ($N_{cut}=10^{11}~\mbox{cm}^{-2}$ 
  for HVC Models B through F and $10^{10}~\mbox{cm}^{-2}$ 
  for HVC Models A and G).
  \label{hvc_si4_comp_obs_fig}}
\end{figure}

\begin{figure}
  \centering
  \epsscale{0.7}
  \plotone{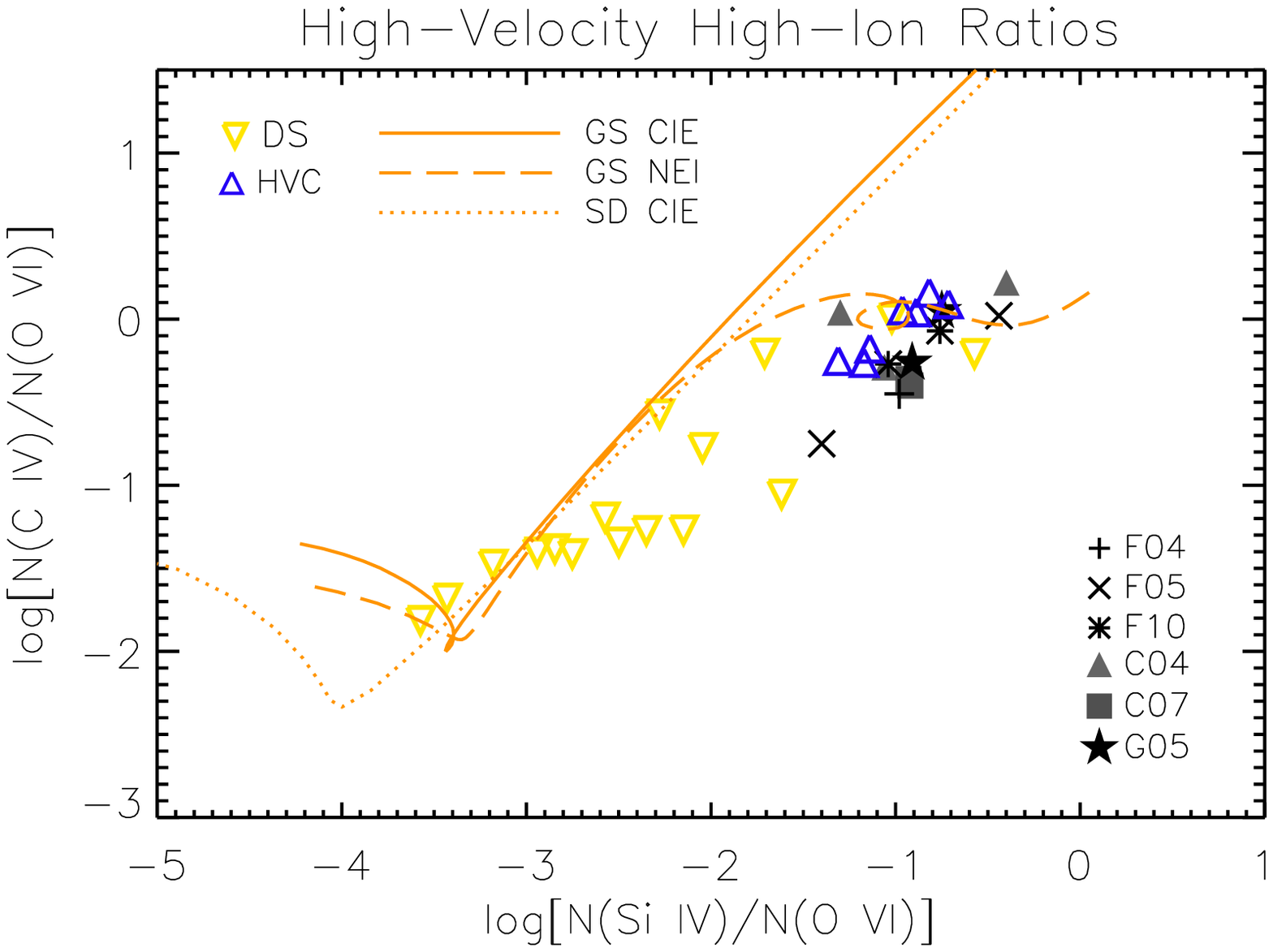}
  \caption{
    \footnotesize{
    N(Si~IV)/N(O~VI) versus N(C~IV)/N(O~VI) in logarithmic scale: 
    comparison between predictions from various models and observations
    for high-velocity ions. 
    Model predictions are presented with colored symbols and lines 
    (whose labels are shown on the upper-left corner) 
    and observations are displayed with black and gray symbols 
    (whose labels are shown on the lower-right corner). 
    Predictions are from the following models:
    shock heating from  
    \citet[][DS: yellow upside-down triangles from their Tables 2A, 2B, 2C, 
    and 2D]{DopitaSutherland1996ApJS}, our HVC models from high-velocity 
    high-ions (blue triangles from mean values in Table \ref{HVC_colm_ratio_T}), 
    and radiative cooling model of hot gas from 
    \citet[][their collisional ionization equilibrium predictions are from 
    their Table 2A, denoted by GS CIE, and traced by a solid orange line, and their 
    isochoric non-equilibrium ionization model predictions are from their Table 2B, 
    denoted by GS NEI, and traced by a dashed orange line]{GnatSternberg2007ApJS} and 
    \citet[][their collisional ionization equilibrium predictions are from their
    Table 5, denoted by SD CIE, and traced by a dotted orange line]
    {SutherlandDopita1993ApJS}. Note that the models indicated by 
    orange lines are more relevant to the low-velocity results shown in 
    Figure \ref{si4o6_c4o6_low_fig}, but we include these models here for 
    convenient comparison between the low- and high-velocity results. 
    Note again that before presenting any of these model predictions, 
    we have converted the assumed abundances to those of \citet{Wilmsetal2000ApJ}. 
    Observations are from the following references: 
    \citet[][F04: $+$ symbols from their Table 4 including the results of detected 
    ion column densities only]{Foxetal2004ApJ},  
    \citet[][F05: $\times$ symbols from their Table 5 including the results of detected
    ion column densities only]{Foxetal2005ApJ},  
    \citet[][F10: $\ast$ symbols from their Tables 1 and 2 including the results of 
    detected ion column densities only]{Foxetal2010ApJ}, 
    \citet[][C04: gray filled triangles from their Tables 2 and 4 including 
    the results of detected ion column densities only]{Collinsetal2004ApJ}, 
    \citet[][C07: gray filled squares from their Tables 4 and 7 including the results 
    of strong line only of the doublet]{Collinsetal2007ApJ}, and 
    \citet[][G05: dark filled stars from their Table 5 including the high-velocity 
    results only, i.e., $v_{LSR} \approx +100$ and $+184$ km $\mbox{s}^{-1}$]
    {Gangulyetal2005ApJS}. 
    }
    \label{si4o6_c4o6_high_fig}
  }
\end{figure}

\begin{figure}
\centering
\epsscale{0.7}
\plotone{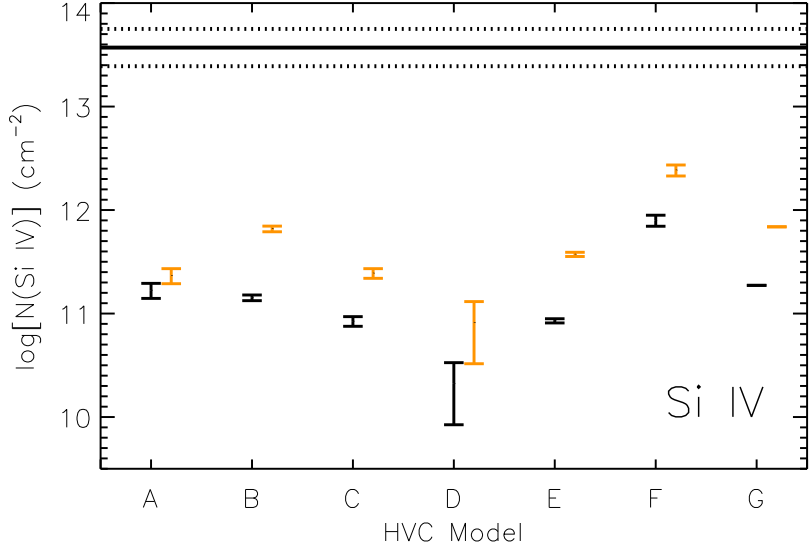}
\caption{Comparison between the average low-velocity Si~IV column density 
  observed on halo sightlines and the predicted average for the halo
  that results from a population of model HVCs. 
  The predicted values from each model are indicated by bars. 
  The black bars are 
  predicted column densities that were obtained from the mass
  integrations to the ends of simulations (columns [2] and [3] in Table
  \ref{low_vel_si4_halo_T}) and the orange bars are the predicted column
  densities that effectively include the mass integrations between 
  the ends of the simulations and the ends 
  of the clouds' lifetimes (columns [5] and [6] in Table
  \ref{low_vel_si4_halo_T}). The top and bottom ends 
  correspond to `Domain+Escaped' and `Domain only',
  respectively, in Table \ref{low_vel_si4_halo_T}. 
  The column densities shown in this figure were 
  calculated with an HVC infall rate 
  $\dot{\mathcal{M}}^{\mathrm{H}}_{\mathrm{HVC}}=0.5 M_{\sun}~\mbox{yr}^{-1}$
  as in Table \ref{low_vel_si4_halo_T}. 
  When the infall rate increases to a factor 2 larger value  
  $\dot{\mathcal{M}}^{\mathrm{H}}_{\mathrm{HVC}}=1.0 M_{\sun}~\mbox{yr}^{-1}$, 
  the column densities also increase by a factor of 2, i.e., both 
  black and orange bars are shifted upwards by a factor of 2.
  The observed values are from \citet{Wakkeretal2012ApJ}. 
  The average of integrated column 
  densities of \ion{Si}{4} along the 32 extra-galactic sightlines 
  from their Figure 6 (top panel for the case of Si~IV) is 
  indicated by the horizontal solid line and its error ranges 
  are shown as dotted lines. 
  \label{si4_halo_fig}}
\end{figure}

\begin{figure}
  \centering
  \epsscale{0.7}
  \plotone{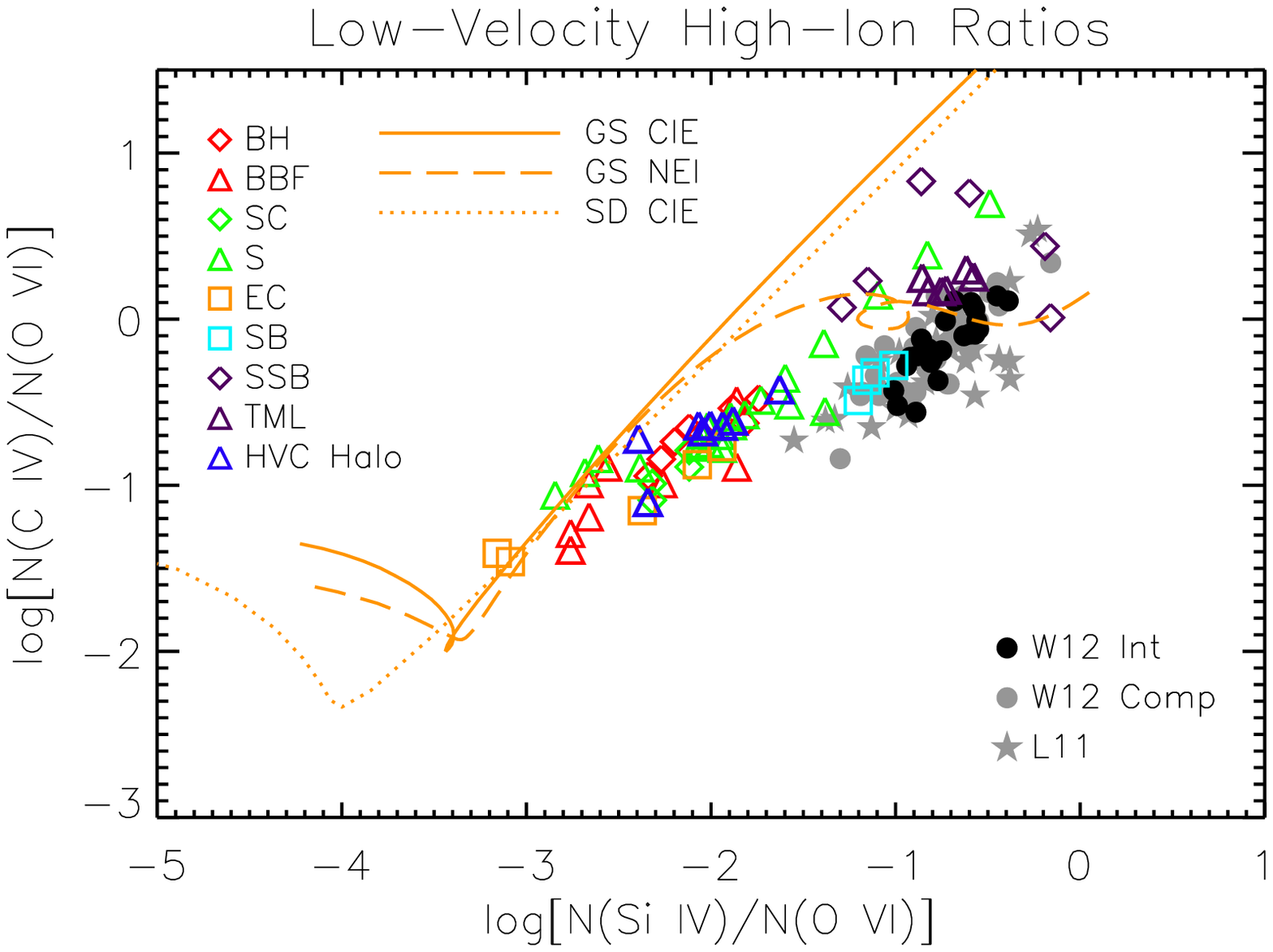}
  \caption{
    \footnotesize{
    N(Si~IV)/N(O~VI) versus N(C~IV)/N(O~VI) in logarithmic scale: 
    comparison between predictions from various models and observations
    for low-velocity ions. 
    Model predictions are presented with colored symbols and lines 
    whose labels are on the upper-left corner 
    and observations are displayed with black and gray symbols 
    whose labels are on the lower-right corner. 
    Predictions are from the following models: thermal conduction from 
    \citet[][BH: red diamonds from their Table 2]{BoehringerHartquist1987MNRAS}
    and \citet[][BBF: red triangles estimated from their Figure 6]
    {Borkowskietal1990ApJ}, 
    supernova remnants from \citet[][SC: green diamonds estimated 
    from their Figure 7]{SlavinCox1992ApJ}
    and \citet[][S: green triangles obtained from the results at 22 epochs 
    of simulation number 103]{Shelton2006ApJ}, 
    radiative cooling of hot gas from 
    \citet[][EC: orange squares from their Table 1]{EdgarChevalier1986ApJL}, 
    \citet[][their collisional ionization equilibrium predictions are from 
    their Table 2A, denoted by GS CIE, and traced by a solid orange line, and their 
    isochoric non-equilibrium ionization model predictions are from their Table 2B, 
    denoted by GS NEI, and traced by a dashed orange line]{GnatSternberg2007ApJS}, and 
    \citet[][their collisional ionization equilibrium predictions are from their
    Table 5, denoted by SD CIE, and traced by a dotted orange line]
    {SutherlandDopita1993ApJS}, 
    galactic fountains from \citet[][SB: cyan squares from their Table 1]
    {ShapiroBenjamin1993sfgi.conf}, and turbulent mixing from 
    \citet[][SSB: purple diamonds from their Table 4A]{Slavinetal1993ApJ} and 
    our TML models (TML: purple triangles from mean values 
    in Table \ref{TLM_colm_ratio_T}). 
    Ratios of low-velocity high ions predicted from our HVC models are represented by 
    blue triangles (HVC Halo). These ratios are calculated from 
    Table 2 of HKS12, where average column densities of \ion{C}{4} and \ion{O}{6} 
    were presented, and Table \ref{low_vel_si4_halo_T}. Among four average column 
    densities in these tables, we take only two of the column densities that include 
    the effect of the cloud's lifetime evolution. 
    The ratios are calculated after taking the average of the two $\beta$-corrected 
    column densities of 'Domain only' and 'Domain + Escaped'.
    Note that before presenting any of these model predictions, we have converted 
    the assumed abundances to those of \citet{Wilmsetal2000ApJ}. 
    Observations are from the following references. 
    Dark and gray filled circles are ratios of integrated and component column 
    column densities, respectively, 
    from \citet{Wakkeretal2012ApJ}. From their Table 2, we extracted 24 ratios of 
    integrated column densities and 41 ratios of component column densities only
    for the detected cases. 
    Gray filled stars are from \citet{Lehneretal2011ApJ} who observed 
    column densities of low-velocity high ions along sightlines to 38 
    stars in the Milky Way disk. From their Table 4, we extracted 29 ratios of 
    detected column densities.
    }
    \label{si4o6_c4o6_low_fig}
  }
\end{figure}

We compared the \ion{Si}{4} column densities extracted from the 
simulations with observations. 
The observational data can be categorized into high velocity 
($|v_{LSR}| \ga 90~\mbox{km}~\mbox{s}^{-1}$)
or low velocity ($|v_{LSR}| \lesssim 30~\mbox{km}~\mbox{s}^{-1}$)
regimes but, for convenience, we combine the low and intermediate 
velocity ranges to a single range encompassing 
$|v_{LSR}| \lesssim 90~\mbox{km}~\mbox{s}^{-1}$ and refer to 
it as low velocity. 
We compare the observed
high velocity column densities with the predicted column densities
from high velocity material in our HVC simulations and we compare
the observed low velocity column densities with those of the decelerated
gas in our HVC models and with the column densities on sightlines
that are perpendicular to the direction of motion in our TML simulations.
We found that the predictions for the high-velocity \ion{Si}{4} 
from our HVC models are in good agreement with the observations. 
Although the column densities of high-velocity 
\ion{Si}{4} calculated along the chosen lines of sight 
from our HVC models vary widely from 
model to model (from $< 10^{11}~\mbox{cm}^{-2}$ 
to a few times $10^{13}~\mbox{cm}^{-2}$), there is overlap 
between the predicted high-velocity column densities and those 
observed \citep[up to $3.2 \times 10^{13} ~\mbox{cm}^{-2}$,][]
{Shulletal2009ApJ}. 
However, when our predictions are
time and space averaged and normalized by the observed population of
HVCs, the resulting quantity of high velocity \ion{Si}{4} is less than
that observed (Section \ref{compare_S_highvel} and 
Figure \ref{hvc_si4_comp_obs_fig}). 
We also compared the ratios of column densities of high-velocity 
high ions, N(\ion{Si}{4})/N(\ion{O}{6}) against 
N(\ion{C}{4})/N(\ion{O}{6}) because comparing the ratios of 
column densities of different ions provides a useful diagnostic to constrain models. 
We found that our HVC models also do a better job of predicting these 
ratios than a general shock heating model (Figure \ref{si4o6_c4o6_high_fig}).

Most of our TML models 
predict low-velocity N(\ion{Si}{4})$\sim$$10^{12}~\mbox{cm}^{-2}$ 
(the exception is TML Model~C, which predicts $\sim 10^{11}~\mbox{cm}^{-12}$) 
so a few tens of layers would be
required if TMLs alone were to explain the typical column density of
low-velocity \ion{Si}{4} observed in the Milky Way's halo. Given the scale
height of \ion{Si}{4}, it may be difficult to fit the required number of
layers along the line of sight.
We also calculated the average column 
densities of low-velocity \ion{Si}{4} predicted from each of our HVC models. 
We found that our individual HVC model clouds produce less decelerated 
\ion{Si}{4} than is actually observed on individual sightlines through the halo.

We also compared the ratios between low-velocity 
N(\ion{Si}{4}), N(\ion{C}{4}), and N(\ion{O}{6}) that are predicted from 
various models including our TML and HVC models and the ratios 
garnered from observations. 
We found that our TML models predict well the ratios of low-velocity ions. 
However, our HVC models underpredict 
the ratios of both N(\ion{Si}{4})/N(\ion{O}{6}) and N(\ion{C}{4})/N(\ion{O}{6}) 
for the low-velocity ions because the relative amounts of \ion{Si}{4} and \ion{C}{4} 
decrease more quickly than those of \ion{O}{6} in the decelerated mixed material. 
Some other models such as the galactic fountain, 
isochoric radiatively cooling, non-equilibrium ionization gas 
in the few times $10^4$ K temperature regime, 
and the late stages of supernova remnant
evolution are also in agreement with the low-velocity
high ions. 
As other researchers did \citep[e.g.,][]{Wakkeretal2012ApJ}, we pointed out the fact 
that all these ``good'' models include the effect of non-equilibrium ionization 
(which is also true for the high-velocity high ions), 
so it is essential to include the non-equilibrium ionization process 
in modeling the production of high ions.

We constructed a model for the low-velocity \ion{Si}{4} 
like the model for the low-velocity \ion{C}{4}, \ion{N}{5}, and \ion{O}{6} in HKS12, 
in which an ensemble of HVCs passes through 
the halo, shedding gas that mixes with halo gas, and the mixed, 
high-ion-rich gas decelerates to the speed of the halo. 
This model explains a much
smaller fraction of the observed \ion{Si}{4} than of the observed \ion{O}{6} 
(e.g., 2\% versus 35\% for HVC Model~B).
We then considered the same composite model as in HKS12, in which 
HVCs are one of four sources of low-velocity high ions in the halo (the other 
sources being extraplanar SNRs, galactic fountains, and photoionization 
by an external radiation field). 
By design, the composite model accounts for all of the observed
low-velocity \ion{O}{6}. HKS12 showed that it can also account for most of the
observed low-velocity \ion{C}{4}. However, this model accounts for less than
half of the observed low-velocity \ion{Si}{4}. 

This shortfall in \ion{Si}{4} might be avoided if the ionizing radiation
were to have a softer spectrum than that used in the photoionized
model component. In that case, photoionization would preferentially
produce \ion{Si}{4}.

\acknowledgments

We are grateful to the anonymous referee for his/her helpful 
comments that improved the original manuscript. 
The FLASH code used in this work was in part developed 
by the DOE-supported ASC/Alliance Center for Astrophysical 
Thermonuclear Flashes at the University of Chicago. 
The simulations were performed at the Research Computing Center (RCC) 
of the University of Georgia. 
This work was supported through grant NNX09AD13G through the 
NASA ATPF program. KK was also supported by Basic Science 
Research Program through the National Research Foundation of 
Korea (NRF) funded by the Ministry of Science, ICT, 
and Future Planning (NRF-2014R1A1A1006050).

\bibliography{apj-jour,ref_colm_dens_Si}

\clearpage

\clearpage

\clearpage

\clearpage

\clearpage

\clearpage

\clearpage

\clearpage

\clearpage

\clearpage

\clearpage

\clearpage

\begin{deluxetable}{cccccccccc}

\tablewidth{0pt}
\tabletypesize{\scriptsize}
\tablecaption{TML Models \label{TMLmodels_T} }
\tablecolumns{10}

\tablehead{
\colhead{} &  
\multicolumn{2}{c}{Domain} &
\multicolumn{3}{c}{Hot Gas} &
\colhead{Initial Interface} &
\multicolumn{3}{c}{Cool Gas} \\
\colhead{} &
\multicolumn{2}{c}{----------------------} &
\multicolumn{3}{c}{--------------------------------------} &
\colhead{} &
\multicolumn{3}{c}{-----------------------------------} \\
\colhead{} &
\colhead{$x$} &
\colhead{$y$} &
\colhead{$n_H$} &
\colhead{T} &
\colhead{$v_x$} &
\colhead{$y=f(x)$} &
\colhead{$n_H$} &
\colhead{T} &
\colhead{$v_x$} \\
\colhead{Model} & \colhead{(pc)} & \colhead{(pc)} & 
\colhead{($\mbox{cm}^{-3}$)} & \colhead{(K)} & 
\colhead{($\mbox{km}~\mbox{s}^{-1}$)} & \colhead{} & 
\colhead{($\mbox{cm}^{-3}$)} & \colhead{(K)} & 
\colhead{($\mbox{km}~\mbox{s}^{-1}$)}
}

\startdata
TML A \tablenotemark{a} & [0, 100] & [-250, 50] & $10^{-4}$ & $10^6$ &
0 & $y=(2.5\mbox{pc})~\sin (\frac{2 \pi x}{100\mbox{pc}})$ & 0.1 &
$10^3$ & 100 \\
TML B \tablenotemark{b} & [0, 100] & [-250, 50] & 
$10^{-4}$ & $10^6$ &
0 & $y=(2.5\mbox{pc})~\sin (\frac{2 \pi x}{100\mbox{pc}})$ & 0.1 &
$10^3$ & 100 \\
TML C & [0, 10] \tablenotemark{c} & [-25, 5] & $10^{-4}$ & $10^6$ &
0 & $y=(0.25\mbox{pc})~\sin (\frac{2 \pi x}{10\mbox{pc}})$ & 0.1 &
$10^3$ & 100 \\
TML D & [0, 100] & [-250, 50] & $10^{-4}$ & $10^6$ &
0 & $y=(2.5\mbox{pc})~\sin (\frac{2 \pi x}{100\mbox{pc}})$ & 0.1 &
$10^3$ & 50 \tablenotemark{d} \\
TML E & [0, 100] & [-250, 50] & $10^{-4}$ & $10^6$ & 0 & 
$y=(5.0\mbox{pc})~\sin (\frac{2 \pi x}{100\mbox{pc}})$
\tablenotemark{e} & 0.1 & $10^3$ & 100 \\
TML F & [0, 100] & [-250, 50] &
$\frac{1}{3}\times10^{-4}$ & $3\times10^6$ \tablenotemark{f} &
0 & $y=(2.5\mbox{pc})~\sin (\frac{2 \pi x}{100\mbox{pc}})$ & 0.1 &
$10^3$ & 100 
\enddata

\tablecomments{Model parameters for TML simulations. 
This table is reproduced from Table 1 of KS10.}

\tablenotetext{a}{Reference TML simulation}
\tablenotetext{b}{TML B is a higher spatial resolution model by a factor of 2
  than TML A.}
\tablenotetext{c}{The computational domain is 1/10 of TML A.}
\tablenotetext{d}{Cool gas has half initial speed of TML A.}
\tablenotetext{e}{Amplitude of initial interface between hot and cool
  gas is twice that of TML A.}
\tablenotetext{f}{The temperature of the hot gas is three times larger than TML A.}

\end{deluxetable}

\clearpage

\begin{deluxetable}{cccccccccc}

\tablewidth{0pt}
\tabletypesize{\scriptsize}
\tablecaption{HVC Models \label{HVCmodels_T} }
\tablecolumns{10}

\tablehead{
\colhead{} & 
\multicolumn{2}{c}{Domain} &
\multicolumn{2}{c}{Ambient Medium} &
\multicolumn{5}{c}{Cloud} \\
\colhead{} &
\multicolumn{2}{c}{-------------------------} &
\multicolumn{2}{c}{-----------------------------} &
\multicolumn{5}{c}{------------------------------------------------------------------------------} \\
\colhead{} &
\colhead{$r$} &
\colhead{$z$} &
\colhead{$n_H$} &
\colhead{T} &
\colhead{Radius \tablenotemark{a}} & 
\colhead{$v_{z,cl}$ \tablenotemark{b}} & 
\colhead{$n_{H,ctr}$ \tablenotemark{c}} &
\colhead{$M_{init,T}$ \tablenotemark{d}} & 
\colhead{$M_{init,v}$ \tablenotemark{e}} \\
\colhead{Model} &
\colhead{(pc)} &
\colhead{(pc)} &
\colhead{($\mbox{cm}^{-3}$)} &
\colhead{(K)} &
\colhead{(pc)} &
\colhead{$\mbox{km}~\mbox{s}^{-1}$} &
\colhead{$\mbox{cm}^{-3}$} &
\colhead{$M_{\sun}$} & 
\colhead{$M_{\sun}$} 
}

\startdata
HVC A & [0,300] & [-200,600] & $1.0\times10^{-4}$ & $10^6$ &
20 & -100 & 0.1 & 120 & 130 \\
HVC B & [0,1200] & [-400,2800] & $1.0\times10^{-4}$ & $10^6$ &
150 & -100 & 0.1 & $4.9\times 10^4$ & $5.1\times 10^4$ \\
HVC C & [0,1200] & [-400,2800] & $1.0\times10^{-4}$ & $10^6$ &
150 & -150 & 0.1 & $4.9\times 10^4$ & $5.1\times 10^4$ \\
HVC D & [0,1200] & [-400,4400] & $1.0\times10^{-4}$ & $10^6$ &
150 & -300 & 0.1 & $4.9\times 10^4$ & $5.1\times 10^4$ \\
HVC E & [0,1200] & [-400,2800] & $1.0\times10^{-4}$ & $10^6$ &
150 & -150 & 0.1 & $4.9\times 10^4$ & $4.9\times 10^4$ \\
HVC F & [0,2400] & [-800,8800] & $1.0\times10^{-4}$ & $10^6$ & 
300 & -100 & 0.1 & $4.0\times 10^5$ & $4.2\times 10^5$ \\
HVC G & [0,1200] & [-400,2800] & $1.0\times10^{-5}$ & $10^6$ & 
150 & -100 & 0.01 & $4.9\times 10^3$ & $5.1\times 10^3$ 
\enddata

\tablecomments{Model parameters for HVC simulations. 
This table is reproduced from Table 1 of KHS11 
with additional information on the domain size.}

\tablenotetext{a}{Approximate radius of 
  the model cloud in HVC simulations except HVC~E, which is a uniform
  density cloud with an exact radius. The HVC model clouds have radial 
  density profiles. See Figure 1 of KHS11 for the 
  detailed density profile of each HVC model cloud.}
\tablenotetext{b}{Initial velocity of
  the cloud along the $z$-direction measured in the observer's
  frame.}
\tablenotetext{c}{Initial hydrogen number density of the cloud at its center.
  Note that inially all the HVC  models have the same cloud temperatures 
  at cloud centers ($T=10^3~\mbox{K}$) and their radial temperature profiles 
  are determined by the cloud density profiles with 
  the condition that the initial pressure is the same in the entire region 
  throughout both the cloud and the ambient medium.}
\tablenotetext{d}{Initial mass of cloud having a 
  temperature $T < 10^4~\mbox{K}$.}
\tablenotetext{e}{Initial mass of cloud moving
  with $v_{z,cl}$. Note that all cloud material with a hydrogen number 
  density greater than 5 times the ambient medium's hydrogen number density
  moves initially at speed $v_{z,cl}$ relative to the ambient medium.}

\end{deluxetable}

\clearpage

\begin{deluxetable}{ccccccccc}

\tablewidth{0pt}
\tabletypesize{\footnotesize}
\tablecaption{Column Density Ratios from TML Models \label{TLM_colm_ratio_T} }
\tablecolumns{9}

\tablehead{
\colhead{} & 
\multicolumn{4}{c}{N(Si~IV) / N(C~IV)} &
\multicolumn{4}{c}{N(Si~IV) / N(O~VI)} \\ 
\colhead{Model} & 
\multicolumn{4}{c}{--------------------------------------------------} &
\multicolumn{4}{c}{--------------------------------------------------} \\
\colhead{} & 
\colhead{~mean} & \colhead{median} &
\colhead{$\sigma$ \tablenotemark{a}} & 
\colhead{[min, max]} & \colhead{~mean} & \colhead{median} &  
\colhead{$\sigma$ \tablenotemark{a}} & 
\colhead{[min, max]} 
}

\startdata
TML A \tablenotemark{b} & 0.12  &   0.12 &   0.0081  &  [0.10, 0.13] &
0.17 &  0.18 &    0.038  & [0.094, 0.24] \\
TML B \tablenotemark{b} & 0.10 &  0.10 &    0.0063 & [0.084, 0.12] &
0.16 &  0.15 &  0.040 & [0.082, 0.23] \\
TML C \tablenotemark{c} & 0.082  & 0.082  &   0.0056  & [0.072, 0.089] & 
0.14 &   0.14 & 0.012  & [0.12, 0.15] \\
TML D \tablenotemark{d} & 0.14  & 0.14  &  0.015 &  [0.10, 0.17] &
0.23 &   0.24 &  0.050 &  [0.15, 0.32] \\
TML E \tablenotemark{b} & 0.12  &  0.12 &  0.0088 & [0.10, 0.14] & 
0.19  &  0.19  &  0.048  & [0.11, 0.29] \\
TML F \tablenotemark{b} & 0.15  &  0.15  &  0.0063  & [0.13, 0.16] &
0.27 &   0.26 &  0.049 &  [0.16, 0.37]
\enddata

\tablenotetext{a}{standard deviation}
\tablenotetext{b}{averaged over $t \in [20, 80]$ Myr}
\tablenotetext{c}{averaged over $t \in [6, 8]$ Myr}
\tablenotetext{d}{averaged over $t \in [30, 80]$ Myr}

\end{deluxetable}

\clearpage

\begin{deluxetable}{ccccccccc}

\tablewidth{0pt}
\tabletypesize{\footnotesize}
\tablecaption{Column Density Ratios from HVC Models \label{HVC_colm_ratio_T} }
\tablecolumns{9}

\tablehead{
\colhead{} & 
\multicolumn{4}{c}{N(Si~IV) / N(C~IV)} &
\multicolumn{4}{c}{N(Si~IV) / N(O~VI)} \\ 
\colhead{Model} & 
\multicolumn{4}{c}{--------------------------------------------------} &
\multicolumn{4}{c}{--------------------------------------------------} \\
\colhead{} & 
\colhead{~mean} & \colhead{median} &
\colhead{$\sigma$ \tablenotemark{a}} & 
\colhead{[min, max]} & \colhead{~mean} & \colhead{median} &  
\colhead{$\sigma$ \tablenotemark{a}} & 
\colhead{[min, max]} 
}

\startdata
HVC A \tablenotemark{b} & 0.20 &   0.16  &      0.12 &  [0.045, 0.60] &
0.26  &  0.19  & 0.18  &  [0.056, 1.0] \\
HVC B \tablenotemark{c} & 0.094  & 0.092 &    0.013  &  [0.071, 0.14] &
0.059  &  0.049  &  0.023 &  [0.026, 0.12] \\
HVC C \tablenotemark{c} & 0.099  &  0.096   &  0.016  &  [0.071, 0.17] &
0.11 &   0.11 &    0.030 &   [0.043, 0.23] \\
HVC D \tablenotemark{c} & 0.11 &   0.10 &    0.013   & [0.082, 0.19] & 
0.17  &   0.15  &   0.065  &   [0.093, 0.35] \\
HVC E \tablenotemark{c} & 0.12  &    0.11   &    0.027  &  [0.086, 0.22] & 
0.16  &     0.13  &     0.061  &  [0.078, 0.36] \\
HVC F \tablenotemark{d} & 0.11  &   0.10   &  0.025  &   [0.076, 0.20] &
0.075  &  0.072   &  0.022    &  [0.034, 0.16] \\
HVC G \tablenotemark{c} & 0.12  &      0.12 &   0.030   &  [0.072, 0.22] &
0.071  & 0.068  &  0.024   &  [0.035, 0.15] 
\enddata

\tablenotetext{a}{standard deviation}
\tablenotetext{b}{calculated over a range of $t \in [10, 60]$ Myr}
\tablenotetext{c}{calculated over a range of $t \in [20, 120]$ Myr}
\tablenotetext{d}{calculated over a range of $t \in [20, 240]$ Myr}

\end{deluxetable}

\clearpage

\begin{deluxetable}{lcc|lcc}

\tablewidth{0pt}
\tabletypesize{\footnotesize}
\tablecaption{Average Column Density of Low-Velocity \ion{Si}{4} in the
  Halo from HVC Simulations \label{low_vel_si4_halo_T}}
\tablecolumns{6}

\tablehead{
\colhead{} &
\multicolumn{2}{c}{~~~Average N(Si~IV)~~~} &
\multicolumn{3}{|c}{Average N(Si~IV)} \\
\colhead{} &
\multicolumn{2}{c}{~~~during the simulations\tablenotemark{a}~~~} &
\multicolumn{3}{|c}{during the cloud's lifetime\tablenotemark{b}} \\
\colhead{} & 
\multicolumn{2}{c}{($10^{11}~\mbox{cm}^{-2}$)} &
\multicolumn{3}{|c}{($10^{11}~\mbox{cm}^{-2}$)} \\
\cline{2-6} & \colhead{~~~~Domain} & \colhead{Domain +} &
\multicolumn{1}{|c}{} & \colhead{Domain} & \colhead{Domain +} \\
\colhead{Model} & \colhead{~~~~only\tablenotemark{c}} & 
\colhead{Escaped\tablenotemark{d}} &
\multicolumn{1}{|l}{~~~$\beta_{HVC}$\tablenotemark{e}~~~} &
\colhead{only\tablenotemark{c}} & \colhead{Escaped\tablenotemark{d}} \\
\colhead{(1)} & \colhead{(2)} & \colhead{(3)} & \colhead{(4)} & 
\colhead{(5)} & \colhead{(6)} 
}

\startdata
HVC A & ~~1.4    &  2.0   &  ~~~0.721  &  1.9   &  2.7  \\
HVC B & ~~1.3    &  1.5   &  ~~~0.216  &  6.2   &  7.0  \\
HVC C & ~~0.75   &  0.93  &  ~~~0.344  &  2.2   &  2.7  \\
HVC D & ~~0.084  &  0.34  &  ~~~0.257  &  0.33  &  1.3 \\
HVC E & ~~0.81   &  0.89  &  ~~~0.228  &  3.6   &  3.9  \\
HVC F & ~~7.0    &  8.9   &  ~~~0.327  &  21    &  27   \\
HVC G & ~~1.9    &  1.9   &  ~~~0.272  &  6.9   &  6.9 
\enddata


\tablecomments{ 
The column densities presented in this table are calculated with 
the same HVC infall rate 
$\dot{\mathcal{M}}^{\mathrm{H}}_{\mathrm{HVC}}=0.5 M_{\sun}~\mbox{yr}^{-1}$ 
as in HKS12. 
When the infall rate increases to a factor 2 larger value  
$\dot{\mathcal{M}}^{\mathrm{H}}_{\mathrm{HVC}}=1.0 M_{\sun}~\mbox{yr}^{-1}$, 
the column densities also increase by a factor of 2. 
}

\tablenotetext{a}{Average N(Si~IV) in the halo due to low-velocity 
  Si~IV produced during the simulation times: $60$ Myrs for HVC Model A, 
  $120$ Myrs for HVC Models B, C, D, E, and G, and $240$ Myrs 
  for HVC Model F}

\tablenotetext{b}{Average N(Si~IV) in the halo due to low-velocity 
  Si~IV produced throughout the cloud's lifetime. 
  The column densities in columns (5) and (6) 
  are obtained by dividing the column densities in columns (2) and (3) 
  by $\beta_{HVC}$ in column (4), respectively, in order to account for 
  the contributions made by ions after the simulations terminated.}

\tablenotetext{c}{Column densities are calculated from Si~IV ions 
  that are counted in the simulation domain.}

\tablenotetext{d}{Column densities are calculated from Si~IV ions 
  in the simulation domain plus Si~IV ions that have escaped through 
  the top boundary of the simulation domain.}

\tablenotetext{e}{
  The fraction of the original HVC H~I mass that has been `removed' 
  due to either ablation or ionization by the 
  end of the simulation. 
}

\end{deluxetable}

\clearpage

\begin{deluxetable}{lcccc}

\tablewidth{0pt}
\tabletypesize{\scriptsize}
\tablecaption{Composite Model of Low-Velocity High Ions
in the Milky Way's Halo\label{composite_T}}
\tablecolumns{5}

\tablehead{
\colhead{} & \colhead{Si~IV} & \colhead{C~IV} &
\colhead{N~V} & \colhead{O~VI}
}

\startdata
~~~~~Observed\tablenotemark{a} ($10^{13}~\mbox{cm}^{-2}$) &
2.9  &  11.3  &  2.5  &  16.2  \\
(1) HVCs\tablenotemark{b} & 
$2\%$ &  $12\%$  &  $22\%$  &  $35\%$  \\
 & ($5\%$) & ($23\%$) & ($44\%$) & ($70\%$) \\
(2) Extraplanar SNRs\tablenotemark{c} &
... & $4\%$ & $8\%$ & $10\%$  \\
(3) Galactic fountains\tablenotemark{d} &
$19-30\%$ & $26-41\%$ & $20-30\%$ & $55\%$ \\
 & ($7-11\%$) & ($10-15\%)$ & ($7-11\%$) & ($20\%$) \\
(4) Photoionization by &  &  &  & \\
~~~~~external radiation field\tablenotemark{e} &
$11\%$ & $25\%$ & ... & ... \\
~~~~~Total & $32-43\%$ & $67-82\%$ & $50-60\%$ & 
$100\%$\tablenotemark{f} \\
 & ($23-27\%$) & ($62-67\%$) & ($59-63\%$) & 
($100\%$)\tablenotemark{f}
\enddata

\tablecomments{Each model component's contribution is 
expressed as a percentage of the observed column density 
of Si~IV, C~IV, N~V, and O~VI. 
The main values are estimated using 
the same HVC infall rate 
$\dot{\mathcal{M}}^{\mathrm{H}}_{\mathrm{HVC}}=0.5 M_{\sun}~\mbox{yr}^{-1}$ 
as in HKS12, while 
the values in the parentheses (in the third, sixth, and last rows) 
are estimated with a factor 2 larger 
infall rate of HVCs, i.e., 
$\dot{\mathcal{M}}^{\mathrm{H}}_{\mathrm{HVC}}=1.0 M_{\sun}~\mbox{yr}^{-1}$
in Equation (\ref{eqn1}). See the text for more details.
}
\tablenotetext{a}{
  Average observed column densities with latitude correction ($N \sin|b|$) 
  calculated from Table 2 of \citet{Wakkeretal2012ApJ}. Only the 
  integrated column densities along the detected sightlines are included. 
  The O~VI value is obtained after removing contribution 
  from the Local Bubble \citep[$\sim 7 \times 10^{12}~\mbox{cm}^{-2}$,][]
  {Oegerleetal2005ApJ}. Note that these new values are slight larger 
  than those in Table 3 of HKS12 for Si~IV, C~IV, and O~VI, while the N~V
  value is the same within errors. 
}
\tablenotetext{b}{
  From our HVC Model~B (reference model); the 
  average of the `Domain-only' and `Domain+Escaped' column 
  densities that were calculated taking into account 
  the lifetime evolution of the model cloud (e.g., for 
  Si~IV, the average of column (5) and column (6) for HVC B in 
  Table \ref{low_vel_si4_halo_T}).
}
\tablenotetext{c}{
  From the model with case 1 and drag coefficient 1 in 
  Table 8 of \citet{Shelton2006ApJ}. The column densities were
  re-calculated according to the abundance adjustment
  from \citet[][used in Shelton]{GrevesseAnders1989AIPC}
  to \citet{Wilmsetal2000ApJ}.
}
\tablenotetext{d}{
  From Table 1 of \citet{ShapiroBenjamin1993sfgi.conf}. The column 
  densities were first re-calculated in order to adjust the abundances 
  from \citet[][used in Shapiro \& Benjamin]{Allen1973asqu.book}
  to \citet{Wilmsetal2000ApJ}, and then rescaled by a factor of 
  0.37 to 0.60 so that the composite model would explain 
  all of the observed O~VI. 
}
\tablenotetext{e}{
  From Figures 3(a) and 3(b) in \citet{ItoIkeuchi1988PASJ}. We assume that 
  negligible N~V and O~VI is produced due to
  photoionization in their model. Again, the column densities of 
  Si~IV and C~IV were re-calculated using the abundances of 
  \citet{Wilmsetal2000ApJ}. (We assumed that the cosmic abundances 
  mentioned in \citet{ItoIkeuchi1988PASJ} are the abundances of 
  \citet{Allen1973asqu.book}.)
}
\tablenotetext{f}{
  By design, our composite model reproduces $100\%$ of the O~VI.
}

\end{deluxetable}

\end{document}